\documentclass[11pt,english,aps,aps,amsmath,showpacs,floats]{revtex4}

\usepackage[T1]{fontenc}
\usepackage[latin1]{inputenc}
\usepackage{color}
\usepackage{float}
\usepackage{bm}
\usepackage{amsmath}
\usepackage{amssymb}
\usepackage{graphicx}
\usepackage{esint}

\makeatletter

\providecommand{\tabularnewline}{\\}

\@ifundefined{textcolor}{}
{%
 \definecolor{BLACK}{gray}{0}
 \definecolor{WHITE}{gray}{1}
 \definecolor{RED}{rgb}{1,0,0}
 \definecolor{GREEN}{rgb}{0,1,0}
 \definecolor{BLUE}{rgb}{0,0,1}
 \definecolor{CYAN}{cmyk}{1,0,0,0}
 \definecolor{MAGENTA}{cmyk}{0,1,0,0}
 \definecolor{YELLOW}{cmyk}{0,0,1,0}
 }

\@ifundefined{definecolor}
 {\usepackage{color}}{}
\@ifundefined{definecolor}{\usepackage{color}}{}
\@ifundefined{definecolor}{\usepackage{color}}{}
\@ifundefined{definecolor}{\usepackage{color}}{}
\@ifundefined{definecolor}{\usepackage{color}}{}
\@ifundefined{definecolor}{\usepackage{color}}{}
\@ifundefined{definecolor}{\usepackage{color}}{}
\usepackage{dcolumn}\usepackage{bm}

\usepackage{babel}

\usepackage{babel}

\usepackage{babel}

\usepackage{babel}

\makeatother

\usepackage{babel}
\begin{document}

\title{Unified decoupling scheme for exchange and anisotropy contributions
and temperature-dependent spectral properties of anisotropic spin
systems}

\author{R. Bastardis$^{1}$, U. Atxitia$^{2}$, O. Chubykalo-Fesenko$^{2}$,
and H. Kachkachi$^{1}$}

\affiliation{$^{1}$PROMES-CNRS UPR8521, Université de Perpignan via Domitia,
Technosud, Rambla de la thermodynamique 66100 Perpignan, France}

\affiliation{$^{2}$Instituto de Ciencia de Materiales de Madrid, CSIC, Cantoblanco,
28049 Madrid, Spain}

\date{\today}
\begin{abstract}
We compute the temperature-dependent spin-wave spectrum and the magnetization
for a spin system using the unified decoupling procedure for the high-order
Green's functions for the exchange coupling and anisotropy, both in
the classical and quantum case. Our approach allows us to establish
a clear crossover between quantum-mechanical and classical methods
by developing the classical analog of the quantum Green's function
technique. The results are compared with the classical spectral density
method and numerical modeling based on the stochastic Landau-Lifshitz
equation and the Monte Carlo technique. As far as the critical temperature
is concerned, there is a full agreement between the classical Green's
functions technique and the classical spectral density method. However,
the former method turns out to be more straightforward and more convenient
than the latter because it avoids any \emph{a priori} assumptions
about the system's spectral density. The temperature-dependent exchange
stiffness as a function of magnetization is investigated within different
approaches. 
\end{abstract}


\maketitle

\section{\label{sec:Introduction}Introduction}

Spin systems offer a rich area for fundamental research, always providing
us with new open and challenging issues. In the context of modern
applications, magnetic systems at the nanoscale have opened a huge
laboratory for testing and applying the available methods with the
challenge to adapt them to the constraints of the new area of magnetic
nanotechnology. Indeed, the rapid development of computers has opened
a new trend for the magnetic materials design. Today the large scale
materials modeling is often used as an efficient way to find optimal
material performance in technological applications such as magnetic
recording. The micromagnetic simulations represent now a powerful
tool, especially after the development of publicly available software
codes. To provide reliable predictions, the modeling methods should
be improved with the incorporation of detailed information from microscopic
materials parameters into the macroscopic parameters such as magnetization,
anisotropy or exchange stiffness. An additional problem arises when
the full-fledged well-known approaches have to be extended to finite-size
systems with acute boundary problems.

Furthermore, multiple recent applications require temperature-dependent
macroscopic properties. These important applications include heat-assisted
magnetic recording \cite{rottmayeretal06ieee}, laser-induced magnetization
dynamics \cite{atxitiaetal07apl}, thermally-assisted magnetic random
memories \cite{prejbeanuetal07jpcm} and thermally-assisted domain
wall motion \cite{hinnow11prl}. In this context the multi-scale scheme
where the temperature-dependent macroscopic parameters are previously
calculated numerically or analytically with the aim to use them in
larger scale modeling has been proposed \cite{kazantsevaetal08prb,atxitiaetal10prb}.
The variety of methods, classical and quantum, analytical and numerical,
were developed in the past and can be adjusted today for applications
within this multi-scale modeling framework. It is then necessary to
take stock of the various methods, compare them and establish their
respective limits of applicability. This is a tremendous task that
has to be tackled before one can apply these methods to design new
magnetic materials.

Accordingly, the present work is about a few standard methods used
for investigating the spectrum of spin waves (SW) in magnetic systems
at finite temperature and for arbitrary spin. These are the quantum
Green's function (QGF) technique and its classical limit (CGF), the
classical spectral density (CSD) method, and the purely numerical
methods, i) one that consists in solving the stochastic Landau-Lifshitz
equation (LLE) \cite{brown63pr,lybcha93jap} and ii) the Metropolis
Monte Carlo (MC) method \cite{binher92}. Although these methods exist
in the literature in different and multiple formulations, no systematic
comparison with the aim to establish their agreement and crossover
has been made. One of our objectives here is to compare these methods
and establish the best framework for the calculation of the temperature-dependent
SW spectrum and physical observables such as the magnetization and
susceptibility. For each method we discuss the most reliable implementation
which gives the best agreement with numerical techniques and provide
a clear crossover between the classical and the quantum case. In this
task we have realized that no unified decoupling scheme used to take
into account both the exchange and anisotropy contributions in the
classical case has been given in the literature so far. However, this
is exactly what is required for the purpose of the hierarchical multi-scale
modeling, where classical Heisenberg-like Hamiltonian is parameterized
via \emph{ab initio} calculations and is used to evaluate temperature-dependent
macroscopic properties \cite{kazantsevaetal08prb,atxitiaetal10prb}.
On the other hand, in the future the use of classical systems may
be avoided if direct reliable calculations of macroscopic properties
at the nanoscale based on quantum spin systems are available. This
is why it is important to establish clear connection between quantum
and classical approaches.

It is well known that the Green's function and spectral density methods
involve a decoupling of high-order spin correlations into two-point
correlations. Here we revisit this issue and demonstrate a clear connection
between, on one hand, the classical and quantum approaches, and on
the other, the CGF technique and the CSD method. In the quantum case,
the spin operators satisfy the $SU(2)$ Lie algebra and this implies
that two spin operators commute when they refer to distinct lattice
sites. In particular, the longitudinal and transverse spin fluctuations
are uncorrelated when they refer to two distinct lattice sites and
they are strongly correlated otherwise. However, a decoupling that
may be successful in dealing with the exchange coupling contribution,
at least at low temperature, may turn out to be a bad approximation
for the local (with identical lattice sites) contributions of (on-site)
anisotropy. This is why mean-field theory (MFT), random-phase approximation
(RPA) and the Bogoliubov-Tyablikov approximation (BTA), which assume
that the longitudinal and transverse fluctuations are uncorrelated,
provide a reasonably good approximation for exchange whereas they
provide rather poor results in the presence of anisotropy.

Here we pay a special attention to this issue and considerably clarify
the situation regarding the decoupling scheme that is used for exchange
and anisotropy contributions. More precisely, we provide a unified
decoupling scheme for both exchange and anisotropy contributions,
for classical as well as quantum spins. Then, using this decoupling
we obtain workable (semi-)analytical expressions for the SW dispersion,
the magnetization and the critical temperature, which are supported
by the good agreement with the numerical results of the LLE method
and Monte Carlo (MC) simulations.

In section \ref{sec:Model} we define the generic system we study
using the Dirac-Heisenberg Hamiltonian. In section \ref{sec:Methods},
we discuss the various decoupling schemes used in QGF technique and
show how they are related, and compute the SW dispersion and the magnetization.
For the latter we also provide the analytical asymptotes at low temperature
and near the critical point. Next, we work out the classical limit
of this approach and obtain the corresponding dispersion and magnetization.
For the latter, we pinpoint an interesting connection between the
Callen's (quantum) expression for the magnetization and the MFT-like
expression in terms of the Brillouin function in the quantum case.
Apart from its elegance, this formulation makes it straightforward
to derive the classical limit in terms of the Langevin function. We
then turn to the CSD method and clarify the relevance of the decoupling
scheme when it comes to treat the exchange coupling and anisotropy.
We end this section with a brief account of the LLE and MC methods
and a few expressions and numerical estimates of the critical temperature.
Section \ref{sec:Magnetization-curves} presents the results for the
SW spectrum, the magnetization as a function temperature and field,
and spin stiffness. The paper ends with a conclusion and outlook.
In Appendices \ref{app:QuantumGreenFunctionMethod} we present the
main steps of the QGF for finite temperature, arbitrary spin, and
oblique magnetic field. In appendix \ref{app:CSDMExchAnisDecoup},
we give a detailed demonstration of some expressions used within the
CSD approach. In appendix \ref{app:TC} we give a few expressions
and numerical estimates for the critical temperature.

\section{\label{sec:Model}Model: Hamiltonian and system studied}

We study a spin system of $\mathcal{N}$ atomic spins $\mathbf{S}_{i}=S\mathbf{s}_{i}$,
with $\left|\mathbf{s}_{i}\right|=1$, interacting via a nearest-neighbor
exchange coupling $J_{ij}$. In addition, each spin evolves in a (local)
potential energy that comprises an on-site anisotropy and a Zeeman
contribution. The anisotropy is taken uniaxial with a common easy
axis pointing in the $z$ direction; the magnetic field is applied
in an arbitrary direction $\mathbf{e}_{h}$ so that $\mathbf{H}=H\mathbf{e}_{h}$.
The Hamiltonian of the system then reads

\begin{equation}
\mathcal{H}=-\frac{1}{2}\sum_{\left\langle i,j\right\rangle }J_{ij}\,\mathbf{S}_{i}\cdot\mathbf{S}_{j}-K\sum_{i=1}^{\mathcal{N}}\left(S_{i}^{z}\right)^{2}-\left(g\mu_{B}H\right){\displaystyle {\displaystyle \sum_{i=1}^{\mathcal{N}}}}\mathbf{S}_{i}\cdot\mathbf{e}_{h}.\label{eq:Hamiltonian}
\end{equation}

We consider only box-shaped systems of size $N_{x}\times N_{y}\times N_{z}=\mathcal{N}$
with, \emph{e.g.} a simple cubic (sc) or a body-centered-cubic (bcc)
lattice structure.

\section{\label{sec:Methods}Methods}

We would like to investigate the spectral properties of such systems
using and comparing two groups of methods: i) the (semi-)analytical
methods, namely the classical spectral density method and the classical
or quantum methods of Green's functions (GF) at finite temperature,
ii) the numerical methods that consist either in solving the stochastic
Landau-Lifshitz-Gilbert equation (LLE) in the Langevin approach or
Monte Carlo (MC) simulations.

\subsection{Quantum Green's function approach}

The Green's function approach has been used thoroughly in almost all
areas of physics. For spin systems, this approach allows us to obtain
and investigate all kinds of observables. As compared to spin-wave
theory (SWT), it makes it possible to obtain in a more systematic
way the excitation spectrum at finite temperature for arbitrary atomic
(nominal) spin.

For our present purposes, we re-derive the basic equations involved
in this approach and apply them to the Hamiltonian (\ref{eq:Hamiltonian}).
In the latter the magnetic field is applied in an arbitrary direction
with respect to the (common) anisotropy easy axis and as such, a slight
reformulation of the basic equations is needed with respect to the
equilibrium configuration. In particular, Callen's formula \citep{akhbarpel68}
for the magnetization in the case of arbitrary spin has to be re-derived
in this context. The details of these calculations are given in Appendix
\ref{app:QuantumGreenFunctionMethod}.

We introduce the retarded many-body Green's functions

\begin{align}
\mathcal{G}^{\mu\nu}(i-j,t) & =\mathcal{G}^{\mu\nu}(\mathbf{r}_{i}-\mathbf{r}_{j},t)\equiv\left\langle \left\langle \sigma_{i}^{\mu}(t);\sigma_{j}^{\nu}(0)\right\rangle \right\rangle _{r}=-i\theta(t)\left\langle \left[\sigma_{i}^{\mu}(t),\,\sigma_{j}^{\nu}(0)\right]\right\rangle .\label{eq:SpinGFr}
\end{align}
 where $\bm{\sigma}_{i}$ are the new spin variables obtained after
rotation of the original variables $\mathbf{S}_{i}$ to the system
of coordinates where the $z$-axis coincides with the direction of
the net magnetization {[}see Appendix \ref{app:QuantumGreenFunctionMethod}{]};
$\left\langle \ldots\right\rangle $ denotes the usual thermal average.
Then, one establishes the equations of motion of the GFs $\mathcal{G}_{ij}^{+-}$,
$\mathcal{G}_{ij}^{--}$, $\mathcal{G}_{ij}^{3-}$ whose solution
renders the SW dispersion.

The equation of motion for a GF of a given order in spin operators
generates GFs of higher orders and this leads to an infinite hierarchy
of GFs satisfying an open system of coupled equations. In order to
close this system of equations and solve it (in Fourier space), one
is led to apply a certain scheme for breaking high-order GFs into
lower-order ones, thus adopting a certain approximation of the magnon-magnon
interactions. Finding an adequate scheme for doing so has triggered
many investigations each dealing with a specific situation with a
particular Hamiltonian. Unfortunately, there is no general or systematic
procedure. In fact, the variety of decoupling schemes only reflects
the complexity of dealing with magnon-magnon interactions and thereby
the nonlinear SW effects. In the following section, we present a discussion
of the main decoupling schemes known in the literature and also propose
some improvements that allow for a certain unification thereof.

\subsubsection{Decoupling schemes}

When applying mean-field theory (MFT), random-phase approximation
(RPA), or the Bogoliubov-Tyablikov approximation (BTA), it is implicitly
assumed that the longitudinal and transverse fluctuations are uncorrelated
and this is a valid approximation only when they refer to distinct
sites. Indeed, the idea behind this approximation consists in writing

\begin{equation}
\left\langle \left[AB,\, C\right]\right\rangle \simeq\left\langle A\right\rangle \times\left\langle \left[B,\, C\right]\right\rangle .\label{eq:BTApprox}
\end{equation}
 For spin systems, the factor $\left\langle A\right\rangle $ is usually
the thermal average of $\sigma^{3}$ and thereby is related to the
temperature-dependent magnetization. Hence, in practice one rearranges
the various terms so that $\sigma^{3}$ appears on the left and then
use the approximation (\ref{eq:BTApprox}). However, in the (local)
anisotropy contributions the product factors are at the same site
and thus the longitudinal and transverse fluctuations are correlated,
which turns this kind of decoupling schemes into relatively bad approximations.

In Ref. \citep{devlin71prev} it was argued that one may avoid this
approximation inherent to a decoupling scheme by establishing $2S$
equations of motion for the anisotropy functions. The problem with
this approach, however, is that in practice one has to specify the
spin $S$ thus limiting the calculations to a particular material.
In addition, it is not obvious how to obtain the classical limit from
the final results.

For the exchange coupling, RPA is commonly used with reasonable satisfaction
since the corresponding results for the SW dispersion and thereby
the magnetization compare fairly well with other techniques such as
Monte Carlo (MC) {[}see Ref. \citep{frokun06pr} for a recent review{]},
as long as the magnetization curve at low temperature is concerned.
However, for a more precise estimation of the critical temperature
$T_{C}$, Callen's decoupling scheme turns out to be much more efficient,
though it leads to a self-consistent equation for $T_{C}$ which is
more difficult to tackle analytically. Indeed, it was shown by Tahir-Kheli
and Callen \cite{tahcal64prev,tah76,callen63prev} that the more sophisticated
decoupling scheme

\begin{align}
\left\langle \left\langle \sigma_{i}^{3}(\tau)\sigma_{l}^{+}(\tau);\sigma_{j}^{-}(0)\right\rangle \right\rangle  & \underset{i\neq j}{\simeq}\left\langle \sigma_{i}^{3}\right\rangle \left\langle \left\langle \sigma_{l}^{+}(\tau);\sigma_{j}^{-}\left(0\right)\right\rangle \right\rangle -\left\langle \sigma_{i}^{3}\right\rangle \frac{\left\langle \sigma_{i}^{-}\sigma_{l}^{+}\right\rangle }{2S^{2}}\left\langle \left\langle \sigma_{i}^{+}(\tau);\sigma_{j}^{-}\left(0\right)\right\rangle \right\rangle \label{eq:DecoupExchTahirKheliCallen}
\end{align}
 takes, to some extent, account of magnon-magnon interactions and
renders a nonlinear equation for the magnon dispersion $\omega(\mathbf{k})$,
see below.

For on-site magneto-crystalline anisotropy the simplistic RPA decoupling
leads to poor and even wrong results. In the presence of anisotropy
with typical ratios $K/J$, the Anderson-Callen decoupling scheme,
originally proposed by Anderson and Callen \citep{callen63prev,andcal64prev}
and later generalized by Schwieger et al. \citep{Schwiegeretal05pr}
to a rotated reference frame, turns out to be rather efficient in
producing reasonable results. This is typically of the form

\begin{align}
K\left\langle \left\langle \left(\sigma_{i}^{3}\sigma_{i}^{-}+\sigma_{i}^{-}\sigma_{i}^{3}\right)\left(t\right);\,\sigma_{j}^{-}\left(0\right)\right\rangle \right\rangle  & \approx2\mathcal{K}_{\sigma}\left\langle \sigma_{i}^{3}\right\rangle \times\left\langle \left\langle \sigma_{i}^{-}\left(t\right);\,\sigma_{j}^{-}\left(0\right)\right\rangle \right\rangle ,\label{eq:ACDecoupling}
\end{align}
 with the effective anisotropy factor

\begin{equation}
\mathcal{K}_{\sigma}=K\left[1-\frac{1}{4S^{2}}\left(\left\langle \sigma_{i}^{+}\sigma_{i}^{-}\right\rangle +\left\langle \sigma_{i}^{-}\sigma_{i}^{+}\right\rangle \right)\right].\label{eq:EffectiveAnisotropy}
\end{equation}

The identity

\[
\left\langle \sigma_{i}^{+}\sigma_{i}^{-}\right\rangle +\left\langle \sigma_{i}^{-}\sigma_{i}^{+}\right\rangle =2S(S+1)-2\left\langle \sigma_{i}^{3}\sigma_{i}^{3}\right\rangle ,
\]
 is derived from the quantum-mechanical identities 
\begin{align}
\sigma^{-}\sigma{}^{+} & =S(S+1)-\left(\sigma^{3}\right)^{2}-\sigma^{3},\nonumber \\
\sigma^{3} & =\frac{1}{2}\left(\sigma^{+}\sigma^{-}-\sigma^{-}\sigma^{+}\right).\label{eq:SU(2)Identities}
\end{align}

It is well known that the decoupling (\ref{eq:ACDecoupling}) is valid
for all spin values $S$ and renders good results when compared with
the exact treatment of anisotropy and with quantum MC when $K/J$
is small \citep{frokun06pr} .

Similar to the decoupling in Eq. (\ref{eq:ACDecoupling}), the following
decoupling for anisotropy has been suggested \cite{andcal64prev}
\begin{align*}
\left\langle \left\langle \sigma_{i}^{3}(\tau)\sigma_{i}^{+}(\tau)+\sigma_{i}^{3}(\tau)\sigma_{i}^{+}(\tau);\sigma_{j}^{-}\left(0\right)\right\rangle \right\rangle  & =\left\langle \sigma_{i}^{3}\right\rangle \left[2-\frac{\left\langle \sigma_{i}^{-}\sigma_{i}^{+}\right\rangle +\left\langle \sigma_{i}^{+}\sigma_{i}^{-}\right\rangle }{2S^{2}}\right]\left\langle \left\langle \sigma_{i}^{+}(\tau);\sigma_{j}^{-}\left(0\right)\right\rangle \right\rangle 
\end{align*}

Then, splitting the right-hand side as follows 
\begin{align*}
\left\langle \sigma_{i}^{3}\right\rangle \left[2-\frac{\left\langle \sigma_{i}^{-}\sigma_{i}^{+}\right\rangle +\left\langle \sigma_{i}^{+}\sigma_{i}^{-}\right\rangle }{2S^{2}}\right]\left\langle \left\langle \sigma_{i}^{+}(\tau);\sigma_{j}^{-}\left(0\right)\right\rangle \right\rangle  & =\left\langle \sigma_{i}^{3}\right\rangle \left(1-\frac{\left\langle \sigma_{i}^{+}\sigma_{i}^{-}\right\rangle }{2S^{2}}\right)\left\langle \left\langle \sigma_{i}^{+}(\tau);\sigma_{j}^{-}\left(0\right)\right\rangle \right\rangle \\
 & +\left\langle \sigma_{i}^{3}\right\rangle \left(1-\frac{\left\langle \sigma_{i}^{-}\sigma_{i}^{+}\right\rangle }{2S^{2}}\right)\left\langle \left\langle \sigma_{i}^{+}(\tau);\sigma_{j}^{-}\left(0\right)\right\rangle \right\rangle 
\end{align*}

we may propose the following decoupling

\begin{align}
\left\langle \left\langle \sigma_{i}^{3}(\tau)\sigma_{i}^{+}(\tau);\sigma_{j}^{-}\left(0\right)\right\rangle \right\rangle  & =\left\langle \sigma_{i}^{3}\right\rangle \left[1-\frac{\left\langle \sigma_{i}^{-}\sigma_{i}^{+}\right\rangle }{2S^{2}}\right]\times\left\langle \left\langle \sigma_{i}^{+}(\tau);\sigma_{j}^{-}\left(0\right)\right\rangle \right\rangle .\label{eq:DecoupAnisAC}
\end{align}

Comparing this decoupling for the anisotropy contribution with Eq.
(\ref{eq:DecoupExchTahirKheliCallen}) for the exchange contribution,
we see that the former follows from the latter upon setting in the
latter $l=i$, \emph{i.e.}, restricting the product of spin operators
to the same lattice site. In fact, this is a consequence of the way
$\sigma_{i}^{3}$ is written in powers of $\sigma_{i}^{3}$ and the
products $\sigma_{i}^{\pm}\sigma_{j}^{\mp}$. More precisely, if we
start from the quantum-mechanical identities (\ref{eq:SU(2)Identities})
and then multiply them by $\alpha$ and $1-\alpha$ respectively and
add the resulting equations we obtain 
\begin{align}
\sigma^{3} & =\alpha S(S+1)-\alpha\left(\sigma^{3}\right)^{2}+\left(\frac{1-\alpha}{2}\right)\sigma^{+}\sigma^{-}-\left(\frac{1+\alpha}{2}\right)\sigma^{-}\sigma{}^{+},\label{eq:SzCallenExpansion}
\end{align}
 where $\alpha$ is then determined so as to comply with the limits
at zero temperature and near the critical point \cite{callen63prev}.
This leads to $\alpha=\left\langle \sigma^{3}\right\rangle /2S^{2}$.

Next, we insert the expression (\ref{eq:SzCallenExpansion}) for $\sigma^{3}$
in products of spin operators such as those appearing on the left-hand
side of Eqs. (\ref{eq:DecoupExchTahirKheliCallen}, \ref{eq:DecoupAnisAC})
and use Wick's or RPA-like decoupling to obtain the decoupling (\ref{eq:DecoupExchTahirKheliCallen})
for exchange and (\ref{eq:DecoupAnisAC}) for anisotropy contributions,
respectively. In fact, there exist several other decoupling schemes
in the literature with expressions for $\alpha$ that are polynomials
of different degrees in $m=\left\langle \sigma^{3}\right\rangle /S$.
Namely, $\alpha=0$ corresponds to RPA (or BTA), $\alpha\propto m$
to Callen's decoupling, $\alpha\propto m^{3}$ to the decoupling proposed
by Copeland and Gersch (CG) \cite{copger66prev}, and 
\begin{equation}
\alpha\left(m\right)=\frac{1}{2}\frac{S-1}{S\left(S+1\right)}m+\frac{1}{S\left(S+1\right)}m^{3}\label{eq:SwendsenDecoup}
\end{equation}
 to the decoupling proposed later by Swendsen \cite{swendsen72prb}.

As already discussed, these polynomials with increasing degrees are
approximations to the more rigorous calculation of spin correlations
that consists in computing contributions of high-order of Feynman's
spin diagrams as is done in Refs. \onlinecite{izyskr88, garlut84fep}.
As it will be seen later in section \ref{sec:Magnetization-curves},
the corresponding decoupling yields fairly precise results for the
magnetization and critical temperature.

\subsubsection{Spin-wave dispersion}

Applying for instance the RPA decoupling to a homogeneous ferromagnet,
\emph{i.e.} with $\left\langle \sigma_{i}^{3}\right\rangle =\left\langle \sigma^{3}\right\rangle $,
(see details in Appendix \ref{app:QuantumGreenFunctionMethod}) one
derives the magnon energy with respect to the equilibrium state

\begin{widetext} 
\begin{eqnarray}
E^{2}(\mathbf{k}) & = & \left(\hbar\omega(\mathbf{k})\right)^{2}=\mathcal{A}_{\mathbf{k}}^{2}-\mathcal{B}_{\mathbf{k}}^{2}\label{eq:MagnonEnergy}\\
 & = & \left[\left(g\mu_{B}\right)\left(H^{x}\sin\vartheta+H^{z}\cos\vartheta\right)+\mathcal{K}_{\sigma}\left\langle \sigma^{3}\right\rangle \left(2\cos^{2}\vartheta-\sin^{2}\vartheta\right)+J_{0}\left\langle \sigma^{3}\right\rangle \left(1-\gamma_{\mathbf{k}}\right)\right]^{2}\nonumber \\
 &  & -\left(\mathcal{K}_{\sigma}\right)^{2}\left\langle \sigma^{3}\right\rangle ^{2}\sin^{4}\vartheta.\nonumber 
\end{eqnarray}
 \end{widetext} where {[}see Appendix \ref{app:QuantumGreenFunctionMethod}
for notation{]} 
\begin{equation}
\begin{array}{lll}
\mathcal{A}_{\mathbf{k}} & \equiv & L^{3}+\mathcal{K}_{\sigma}\left\langle \sigma^{3}\right\rangle \left(2\cos^{2}\vartheta-\sin^{2}\vartheta\right)+J_{0}\left\langle \sigma^{3}\right\rangle \left(1-\gamma_{\mathbf{k}}\right),\\
\mathcal{B}_{\mathbf{k}} & \equiv & \mathcal{K}_{\sigma}\left\langle \sigma^{3}\right\rangle \sin^{2}\vartheta
\end{array}\label{eq:EMSystemCoeffs}
\end{equation}
 $J_{0}=J\left(0\right)$ being defined in Eq. (\ref{eq:zeroCompJ}).

This dispersion relation is effectively obtained within the linear
spin-wave theory, because the high-order GFs stemming from exchange
contributions have been decoupled using the RPA which does not take
account of spin correlations or magnon-magnon interactions. Indeed,
following the standard procedure described in the appendix of Ref.
\citep{atxitiaetal10prb} (and references therein) one arrives at
the equation for the dispersion (in the case $\vartheta=0$, \emph{i.e.},
of longitudinal field and $\left\langle \sigma_{i}^{3}\right\rangle =\left\langle S_{i}^{z}\right\rangle $)

\begin{align}
\hbar\omega\left(\mathbf{k}\right)\equiv\hbar\omega_{\mathbf{k}} & =\left(g\mu_{B}\right)H+2\mathcal{K}_{S}\left\langle S^{z}\right\rangle +\left\langle S^{z}\right\rangle \left[J(0)-J(\mathbf{k})\right]+\frac{\left\langle S^{z}\right\rangle }{\mathcal{N}S}\alpha\sum_{\mathbf{p}}\left[J(\mathbf{p})-J(\mathbf{p}-\mathbf{k})\right]\left\langle n_{\mathbf{p}}\right\rangle ,\label{eq:NLMagnonDispGF}
\end{align}
 where $\mathcal{K}_{S}$ reads,

\begin{equation}
\mathcal{K}_{S}=\mathcal{K}_{\sigma}\left(\vartheta=0\right)=K\left[1-\frac{1}{2S^{2}}\left(S(S+1)-\left\langle S^{z}S^{z}\right\rangle \right)\right]\label{eq:EffectiveAnisotropy-1}
\end{equation}
 and $\alpha=m$ for Callen's decoupling. $\left\langle n_{\mathbf{p}}\right\rangle $
is the thermal occupation number given by the magnon Bose-Einstein
distribution 
\begin{equation}
\left\langle n_{\mathbf{p}}\right\rangle =\frac{1}{e^{\beta\hbar\omega_{\mathbf{p}}}-1}\label{eq:OccupationNumb}
\end{equation}
 where $\beta=1/k_{\mathrm{B}}T$.

Then, using translation invariance we see that 
\[
\frac{J(\mathbf{p})-J(\mathbf{p}-\mathbf{k})}{J(0)-J(\mathbf{k})}=\frac{\gamma_{\mathbf{p}}-\gamma_{\mathbf{p}\mathbf{-k}}}{1-\gamma_{\mathbf{k}}}=\gamma_{\mathbf{p}}
\]
 and thereby 
\begin{align}
\hbar\omega_{\mathbf{k}} & =\left(g\mu_{B}\right)H+2\mathcal{K}_{S}\left\langle S^{z}\right\rangle +J_{0}\left\langle S^{z}\right\rangle Q\left(\alpha,\beta\right)\left(1-\gamma_{\mathbf{k}}\right).\label{eq:NLMagnonDispGFgamma}
\end{align}
 Here we have introduced the exchange {}``stiffness coefficient''

\begin{align}
Q(\alpha,\beta) & =1+\frac{\alpha}{\mathcal{N}S}\sum_{\mathbf{p}}\frac{\gamma_{\mathbf{p}}}{e^{\beta\hbar\omega_{\mathbf{p}}}-1}=1+\frac{\alpha}{\mathcal{N}S}\sum_{\mathbf{p}}\gamma_{\mathbf{p}}\left\langle n_{\mathbf{p}}\right\rangle \label{eq:QExpression}
\end{align}
 where $\alpha$ is, as defined earlier, depends on the decoupling
scheme.

\subsubsection{Magnetization}

Now, we turn to compute the magnetization for an arbitrary spin $S$.
In Callen's method {[}see Ref. \citep{akhbarpel68} and references
therein{]} one considers the GF

\begin{equation}
\Xi^{\mu\nu}(i-j,t;\xi)=-i\theta(t)\times\left\langle \left[\sigma_{i}^{\mu}(t),\exp(\xi\sigma_{j}^{z}(0))\;\sigma_{j}^{\nu}(0)\right]\right\rangle .\label{eq:CallenGF}
\end{equation}

Then, replacing the GFs in the system (\ref{eq:GFSystem}) by their
analogs from Eq. (\ref{eq:CallenGF}) we obtain the new system of
EM

\begin{widetext} 
\begin{equation}
\left\{ \begin{array}{ccc}
\left(\omega-\mathcal{A}_{\mathbf{k}}\right)\Xi_{\mathbf{k}}^{+-}+\mathcal{B}_{\mathbf{k}}\Xi_{\mathbf{k}}^{--} & = & \left\langle \left[\sigma_{i}^{+}(0),\exp(\xi\sigma_{i}^{z}(0))\;\sigma_{i}^{-}(0)\right]\right\rangle \equiv\Sigma^{+-}\\
\\
-\mathcal{B}_{\mathbf{k}}\Xi_{\mathbf{k}}^{+-}+\left(\omega+\mathcal{A}_{\mathbf{k}}\right)\Xi_{\mathbf{k}}^{--} & = & 0.
\end{array}\right.\label{eq:CallenSEM}
\end{equation}
 \end{widetext}

Following again Callen's procedure we obtain for the first moment
$\left\langle \sigma^{3}\right\rangle $

\begin{equation}
\left\langle \sigma^{3}\right\rangle =\frac{(S-\Omega)(1+\Omega)^{2S+1}+(S+1+\Omega)\Omega^{2S+1}}{(1+\Omega)^{2S+1}-\Omega^{2S+1}}.\label{eq:CallenMag}
\end{equation}
 where $\Omega$ is the following function of the $1^{\mathrm{st}}$
and $2^{\mathrm{nd}}$ moments $\left\langle \sigma^{3}\right\rangle ,\left\langle \sigma^{3}\sigma^{3}\right\rangle \equiv\mathcal{C}^{33}$,

\begin{equation}
\Omega\left(\left\langle \sigma^{3}\right\rangle ,\mathcal{C}^{33}\right)=\frac{1}{2}\frac{1}{N_{c}}\sum_{\mathbf{k}}\left[\frac{\mathcal{A}_{\mathbf{k}}}{\omega_{\mathbf{k}}}\coth\left(\frac{\beta\hbar\omega_{\mathbf{k}}}{2}\right)-1\right].\label{eq:OmegaK}
\end{equation}
 $N_{c}$ is the number of unit cells in the Bravais lattice of the
ferromagnet and is also the number of allowed wave-vectors in the
Brillouin zone.

Eq. (\ref{eq:CallenMag}), together with (\ref{eq:NLMagnonDispGFgamma})
and (\ref{eq:OmegaK}), constitutes a transcendental equation whose
solution involves several sums (integrals) in Fourier space. In general,
it is a heavy task to solve Eq. (\ref{eq:CallenMag}) especially for
lattices with several sub-lattices. Nonetheless, it was shown in Ref.
\citep{calstr65ssc} that the magnetization $\bar{\sigma}\equiv\left\langle \sigma^{3}\right\rangle $
in Eq. (\ref{eq:CallenMag}) can be recast into the following more
compact form 
\begin{equation}
\bar{\sigma}=SB_{S}\left[SX\right]\label{eq:Callen-Brillouin}
\end{equation}
 where $X$ is defined by 
\begin{equation}
\Omega=\frac{1}{e^{X}-1}\label{eq:MFTExcitDensity}
\end{equation}
 and $B_{S}(x)$ is the Brillouin function (for the quantum spin $S$)
\begin{equation}
B_{S}(x)=\frac{2S+1}{2S}\coth\left[\left(\frac{2S+1}{2S}\right)x\right]-\frac{1}{2S}\coth\left(\frac{x}{2S}\right).\label{eq:Brillouin}
\end{equation}

More generally, it was shown \citep{calcal65prev,calstr65ssc} that
the higher moments $\left\langle \bar{\sigma}^{n}\right\rangle $
can all be expressed in terms of the reduced magnetization $m=\bar{\sigma}/S$
whereby the temperature $T$ and field $H$ enter via $m=m\left(T,H\right)$.
It was argued that this model-independent MFT-like result stems from
the exponential form of the probability density, \emph{i.e.}, $\rho=e^{XS^{z}}/\mathrm{Tr}\, e^{XS^{z}}$.
Indeed, Eq. (\ref{eq:MFTExcitDensity}) expresses the fact that in
MFT all the excitations are degenerate and that one may define the
energy $\varepsilon=X/k_{\mathrm{B}}T$ as the effective energy of
the molecular-field-like excitations with the same occupation number
as the true excitations. On the other hand, we note that Eq. (\ref{eq:Callen-Brillouin})
is also a transcendental equation for $\bar{\sigma}$, similar to
Eq. (\ref{eq:CallenMag}), though much more compact and it readily
yields the classical limit, as will be seen below. In addition, this
establishes the connection to the standard result of MFT.

In order to solve either equation, \emph{i.e.} (\ref{eq:CallenMag})
or (\ref{eq:Callen-Brillouin}), and obtain the magnetization $m\left(T,H\right)$
one has to supplement the latter by a second equation for the correlation
function $\mathcal{C}^{33}=\left\langle \sigma^{3}\sigma^{3}\right\rangle $;
this is obtained by the Callen's procedure (for $\xi=0$, see Eq.
(\ref{eq:CallenGF})) which leads to 
\begin{eqnarray}
\left\langle \sigma_{i}^{-}\sigma_{i}^{+}\right\rangle  & = & 2\bar{\sigma}\times\Omega\left(\bar{\sigma},\mathcal{C}^{33}\right).\label{eq:CorrEq}
\end{eqnarray}
 together with $1^{\mathrm{st}}$ the identity in Eq. (\ref{eq:SU(2)Identities}).

The latter also yields 
\begin{eqnarray}
\left\langle \sigma^{3}\sigma^{3}\right\rangle  & = & S\left(S+1\right)-\bar{\sigma}-\left\langle \sigma^{-}\sigma^{+}\right\rangle =S\left(S+1\right)-\left(1+2\Omega\right)\bar{\sigma}.\label{eq:2ndMoment}
\end{eqnarray}

Finally, the magnetization (in the rotated frame) $\bar{\sigma}$
is given by the solution of the following system of two nonlinear
(coupled) equations 
\begin{equation}
\left\{ \begin{array}{ccc}
\bar{\sigma} & = & SB_{S}\left[S\ln\left(1+\frac{1}{\Omega}\right)\right],\\
\\
\mathcal{C}^{33} & = & S\left(S+1\right)-\left(1+2\Omega\right)\bar{\sigma}.
\end{array}\right.\label{eq:1st2snOrderEqs}
\end{equation}

In general, this system can only be solved numerically as it involves
transcendental equations with several integrals. However, we can establish
a few analytical expressions for the magnetization in the limiting
temperature regions $T\rightarrow0$ and $T\to T_{C}$ and upon restricting
ourselves to a longitudinal magnetic field, i.e. applied along the
direction $\mathbf{e}_{3}$ ($\psi=0$).

In this case the SW dispersion $\hbar\omega_{\mathbf{k}}$ in Eq.
(\ref{eq:MagnonEnergy}) simplifies into

\[
\hbar\omega_{\mathbf{k}}=\mathcal{A}_{\mathbf{k}}=L^{3}+2\mathcal{K}_{\sigma}\bar{\sigma}+J_{0}\bar{\sigma}\left(1-\gamma_{\mathbf{k}}\right)
\]
 and Eq. (\ref{eq:OmegaK}) becomes

\begin{equation}
\Omega\left(\bar{\sigma},\mathcal{C}^{33}\right)=\frac{1}{N_{c}}\sum_{\mathbf{k}}\frac{1}{2}\left[\coth\left(\frac{\beta\hbar\omega_{\mathbf{k}}}{2}\right)-1\right].\label{eq:LAOmega}
\end{equation}

\medskip{}

\textbf{Low temperature asymptote}

At low temperature, the spins are strongly correlated and thereby
the correlation function $\mathcal{C}^{33}$ tends to $S\left(S+1\right)$.
As a consequence, the effective anisotropy obtained from the Anderson-Callen
decoupling scheme simply yields $\mathcal{K}_{S}\longrightarrow K$
{[}see Eq. (\ref{eq:EffectiveAnisotropy-1}){]} so that the system
of equations (\ref{eq:1st2snOrderEqs}) decouples leading to a closed
equation for $\bar{\sigma}$ whose solution then is Eq. (\ref{eq:CallenMag}).
Expanding the latter in terms of $\Omega$ (which becomes small at
low temperature), we find 
\[
\bar{\sigma}\simeq S-\Omega\left(\bar{\sigma}\right).
\]

Moreover, at low temperature only low-energy spin waves are excited
and these are the long-wave length modes. Hence, in the limit of small
wave vectors, we have the dispersion relation 
\begin{equation}
\hbar\omega_{\mathbf{k}}\simeq g\mu_{B}H^{z}+2K\bar{\sigma}+A\bar{\sigma}k^{2},\label{eq:LowTDisp}
\end{equation}
 where $A\equiv J\delta^{2}$, $\delta^{2}\equiv\sum a^{2}J(a)/\sum J(a)$,
$a$ is the lattice parameter and $J(a)$ is the exchange coupling
over the nearest-neighbor bond.

Next, upon expanding $\Omega$ in terms of temperature $T$ (or rather
in $k_{\mathrm{B}}T/J_{0}S$) we obtain 
\begin{equation}
\left\langle \sigma^{3}\right\rangle \simeq S-\left(\frac{3\tau}{2\pi S}\right)^{3/2}Z{}_{3/2}\left[\frac{h+\kappa S}{\tau}\right]\label{eq:LowTMag}
\end{equation}
 where 
\[
Z_{p}\left(x\right)=\sum_{n=1}^{\infty}n^{-p}e^{-nx},
\]
 and $Z_{p}\left(0\right)=\zeta\left(p\right)$ is the well known
Riemann zeta function. We have also introduced the following dimensionless
parameters

\[
\tau\equiv\frac{1}{\beta J_{0}}=\frac{k_{\mathrm{B}}T}{J_{0}},\quad h\equiv\frac{\left(g\mu_{B}\right)H}{J_{0}},\quad\kappa\equiv\frac{2K}{J_{0}}.
\]

Obviously, in the present limit and in zero applied field, one obtains
the well known {}``$3/2$'' Bloch's power law for the thermal decrease
of the magnetization.

\medskip{}

\textbf{Near-critical temperature asymptote ($H=0$)}

Just below the Curie temperature, in the absence of magnetic field,
the mean number of excited quasi-particles and their density are large,
and it is then a reasonable approximation to pass to the continuum
limit. In this case, in Eq. (\ref{eq:LAOmega}) we make the transformation
\[
\frac{1}{N_{c}}\times\sum_{\mathbf{k}}\left(\cdots\right)_{\mathbf{k}}\longrightarrow\frac{1}{N_{c}}\times\frac{V}{\left(2\pi\right)^{3}}\iiint d^{3}k\,\left(\cdots\right)_{\mathbf{k}}=\frac{V}{N_{c}}\iiint\frac{d^{3}k}{\left(2\pi\right)^{3}}\left(\cdots\right)_{\mathbf{k}}=v_{0}\int\frac{d\mathbf{k}}{\left(2\pi\right)^{3}}\left(\cdots\right)_{\mathbf{k}}
\]
 where $v_{0}$ is the volume of the unit cell of the direct lattice.

Next, in this limit the system (\ref{eq:1st2snOrderEqs}) again decouples
and leads to the Callen's expression (\ref{eq:CallenMag}) for the
magnetization, similarly to the low-temperature limit. In addition,
we may write for $C^{33}$

\begin{equation}
C^{33}\simeq\frac{S\left(S+1\right)}{3}\label{eq:HighTCorrelator}
\end{equation}
 where the factor $1/3$ stems from the three dimensional rotational
symmetry ($SO\left(3\right)$) of spins that starts to recover as
the temperature reaches the critical temperature of the ferromagnet.

Consequently, upon inserting in Eq. (\ref{eq:NLMagnonDispGF}) $\mathcal{C}^{33}=\left\langle S^{z}S^{z}\right\rangle $
given by the result above, dropping the nonlinear SW contributions,
and neglecting the second-order terms in $\bar{\sigma}$ we obtain
the dispersion

\begin{align}
\hbar\omega_{\mathbf{k}} & =2K\eta\bar{\sigma}+\bar{\sigma}J_{0}\left(1-\gamma_{\mathbf{k}}\right)=J_{0}\bar{\sigma}\lambda^{-1}\left(1-\lambda\gamma_{\mathbf{k}}\right).\label{eq:NearTcDisp}
\end{align}
 with 
\begin{equation}
\eta\equiv1-\frac{S(S+1)}{3S^{2}},\quad\lambda\equiv\frac{1}{1+\eta\kappa}.\label{eq:NearTcDispParams}
\end{equation}

In addition, $\bar{\sigma}$ is rather small because the density of
SW is large and since $\omega_{\mathbf{k}}$ is proportional to $\bar{\sigma}$,
as is seen in Eq. (\ref{eq:NearTcDisp}), we can expand $\Omega\left(\left\langle \sigma^{3}\right\rangle \right)$
in powers of $\omega_{\mathbf{k}}$ and obtain

\begin{align}
\Omega\left(\bar{\sigma}\right) & =v_{0}\intop\frac{d\mathbf{k}}{\left(2\pi\right)^{3}}\frac{1}{e^{\beta\hbar\omega_{\mathbf{k}}}-1}\simeq v_{0}\intop\frac{d\mathbf{k}}{\left(2\pi\right)^{3}}\left(\frac{1}{\beta\hbar\omega_{\mathbf{k}}}-\frac{1}{2}+\frac{\beta\hbar\omega_{\mathbf{k}}}{12}\right).\label{eq:OmegaNearTc}
\end{align}

Let us now compute these integrals. Using (\ref{eq:NearTcDisp}),
the first contribution reads

\[
v_{0}\intop\frac{d\mathbf{k}}{\left(2\pi\right)^{3}}\frac{1}{\beta\hbar\omega_{\mathbf{k}}}=\frac{\lambda P\left(\lambda\right)}{\bar{\sigma}}\tau.
\]
 where we have introduced the well known lattice Green's function
{[}see Ref. \onlinecite{joy72pt} and references therein{]}

\begin{equation}
P\left(\lambda\right)\equiv v_{0}\intop\frac{d\mathbf{k}}{\left(2\pi\right)^{3}}\frac{1}{1-\lambda\gamma_{\mathbf{k}}}.\label{eq:LatticeGF}
\end{equation}
 Analytical expressions for this integral for various limiting cases
of the parameter $\lambda$ are given in Ref. \onlinecite{joy72pt},
see also Eq. (4.2) in Ref. \onlinecite{gar96prb}. In our case,
from (\ref{eq:NearTcDispParams}) we have $\lambda=\left(1+\eta\kappa\right)^{-1}\simeq1-\eta\kappa$
since $\kappa=2K/J_{0}\ll1$. Hence $\delta\lambda\equiv1-\lambda\ll1$
and according to Refs. \onlinecite{joy72pt, gar96prb} we have 
\begin{equation}
P\left(\lambda\right)\simeq W-c_{0}\left(1-\lambda\right)^{1/2}.\label{eq:LatticeGFExpand}
\end{equation}

$W$ is the Watson integral that evaluates to $1.51639$ for a sc
lattice and to $1.39320$ for a bcc lattice; $c_{0}$ is a lattice-dependent
constant that is equal to $\frac{3}{\pi}\left(\frac{3}{2}\right)^{1/2}\simeq1.16955$
for the sc lattice and to $2^{3/2}/\pi\simeq0.90032$ for the bcc
lattice \cite{joy72pt}. Next, using the fact that for both sc and
bcc lattices \cite{joy72pt} 
\begin{equation}
\intop\frac{d\mathbf{k}}{\left(2\pi\right)^{3}}\left(\gamma_{\mathbf{k}}\right)^{2n+1}=0,\quad n=0,1,2,\ldots\label{eq:PowersGamma}
\end{equation}
 we compute the remaining contributions in (\ref{eq:OmegaNearTc})
and obtain 
\begin{align*}
\Omega\left(\bar{\sigma}\right) & \simeq\frac{\lambda P\left(\lambda\right)\tau}{\bar{\sigma}}-\frac{1}{2}+\frac{1}{12\lambda}\frac{\bar{\sigma}}{\tau}.
\end{align*}

Finally, using this expression in Eq. (\ref{eq:CallenMag}) and expanding
with respect to $\bar{\sigma}$ we obtain the following asymptote
for the magnetization

\begin{align}
\bar{\sigma}^{\textrm{QGF}} & \simeq\frac{2\sqrt{15}\lambda P\left(\lambda\right)\tau}{\sqrt{4S(S+1)+5P(\lambda)-3}}\sqrt{1-\frac{3\lambda P\left(\lambda\right)\tau}{S(S+1)}}.\label{eq:MagQGFNearTc}
\end{align}

This asymptotic expression is plotted in Fig. \ref{fig:m(T)QGF2CGF}
where it favorably compares with the other (exact numerical) magnetization
curves.

Now, in this relatively high temperature regime, magnon-magnon interactions
become relevant. In order to take them into account, we consider the
dispersion in Eq. (\ref{eq:NearTcDisp}) to which we add the last
contribution in Eq. (\ref{eq:NLMagnonDispGFgamma}), \emph{i.e.}

\[
\hbar\omega_{\mathbf{k}}=J_{0}\bar{\sigma}Q(\alpha,\tau)\Lambda^{-1}\left(1-\Lambda\gamma_{\mathbf{k}}\right),
\]
 where 
\begin{equation}
\left\{ \begin{array}{lll}
\Lambda & \equiv & \frac{Q(\alpha,\tau)}{\kappa+Q(\alpha,\tau)},\\
\\
Q(\alpha,\tau) & = & 1+\frac{\alpha}{S^{2}\mathcal{N}}\sum_{\mathbf{p}}\frac{\gamma_{\mathbf{p}}}{e^{\beta\hbar\omega_{\mathbf{p}}}-1}.
\end{array}\right.\label{eq:LQ}
\end{equation}

Upon neglecting the second-order terms in $\bar{\sigma}$ the last
expression leads to the transcendental equation for $Q$

\begin{align}
Q(\alpha,\tau) & \simeq1+\frac{\tau}{Q(\alpha,\tau)}\left[W_{N}-C_{0}\sqrt{1-\frac{Q(\alpha,\tau)}{\kappa+Q(\alpha,\tau)}}-1\right].\label{eq:QphitT}
\end{align}

In the absence of anisotropy one can easily solve the latter and obtain

\begin{align}
Q(\alpha,\tau) & \simeq\frac{1}{2}\left(1+\sqrt{1+4\frac{\alpha}{m}\tau\left(W_{N}-1\right)}\right)\equiv Q_{\mathrm{exch}}(\alpha,\tau).\label{eq:Qexchange}
\end{align}

Since the anisotropy contribution is much smaller than that of exchange
we may seek a solution for $Q\left(\alpha,\tau\right)$ in the form

\[
Q\left(\alpha,\tau\right)\simeq Q_{\mathrm{exch}}(\alpha,\tau)\left(1+\epsilon\right),\qquad\epsilon\equiv\frac{Q_{\mathrm{exch}}}{Q_{\mathrm{anis}}}.
\]
 Then, inserting this in Eq. (\ref{eq:QphitT}) and expanding successively
with respect to $\epsilon$ and then with respect to $\kappa$, we
obtain (to first order)

\begin{align}
Q(\alpha,\tau) & \simeq Q_{\mathrm{exch}}(\alpha,\tau)-\phi\frac{C_{0}\tau}{2Q_{\mathrm{exch}}(\alpha,\tau)-1}\sqrt{\frac{\kappa}{\kappa+Q_{\mathrm{exch}}(\alpha,\tau)}}.\label{eq:QexhangeAniso}
\end{align}

Next, using the same expansion for $\Omega\left(\left\langle \sigma^{3}\right\rangle \right)$,
similar to Eq. (\ref{eq:OmegaNearTc}) we get

\begin{align*}
\Omega\left(\bar{\sigma}\right) & \simeq\frac{\Lambda P\left(\Lambda\right)}{\bar{\sigma}}\tau-\frac{1}{2}+\frac{1}{12\Lambda}\frac{\bar{\sigma}}{\tau}
\end{align*}
 which leads to the following asymptote for the magnetization

\begin{equation}
\bar{\sigma}^{\textrm{QGF}}\simeq\frac{\sqrt{15}\Lambda P(\Lambda)\tau}{Q(\alpha,\tau)\sqrt{S(S+1)+\frac{5P(\Lambda)-3}{4}}}\sqrt{1-\frac{3\Lambda P(\Lambda)\tau}{S(S+1)Q(\alpha,\tau)}}.\label{eq:MagCGFNearTCPhi}
\end{equation}

Note that this expression reduces to that in Eq. (\ref{eq:MagQGFNearTc})
if we set $\alpha=0$ since then $Q(0,\tau)=1$ and $\Lambda=\lambda$,
which corresponds to the RPA decoupling. On the other hand, as we
will see later {[}see Eq. (\ref{eq:QGFTc}){]}, one has to use this
expression instead of (\ref{eq:MagQGFNearTc}) to obtain the critical
temperature. In addition, as far as the magnetization is concerned,
Eq. (\ref{eq:MagCGFNearTCPhi}) renders a more precise profile for
relatively higher temperatures.

\subsection{Classical Green's function approach}

In many situations, the classical approach turns out to be appropriate
for describing the magnetic properties of the system studied. Therefore,
it is worth establishing analogous expressions as in the quantum case
by carefully examining the corresponding decoupling schemes and controlling
the various approximations. Accordingly, in this section we establish
a complete procedure, analogous to the quantum-mechanical one, that
yields the classical SW dispersion and thereby the magnetization.
In particular, we provide the classical analog of Callen's decoupling
scheme, for both exchange and anisotropy contributions.

For this purpose we first set $\vartheta=0$ and return to the spin
variables $\mathbf{S}_{i}$. We then introduce the classical spin
vectors $\mathbf{s}_{i}=\mathbf{S}_{i}/S$ and make the substitutions
$J_{ij}\rightarrow J'_{ij}=S^{2}J_{ij},K\rightarrow K'=S^{2}K,H\rightarrow H'=SH$
in the Hamiltonian (\ref{eq:Hamiltonian}). Next, we define the classical
two-time (retarded) GF

\begin{equation}
G_{ij}(\tau)=\left\langle \left\langle s_{i}^{+}(\tau);s_{j}^{-}(0)\right\rangle \right\rangle =-i\theta(\tau)\mbox{\ensuremath{\left\langle \left\{ s_{i}^{+}(\tau),s_{j}^{-}(0)\right\} \right\rangle }},\label{eq:RetardedCGF}
\end{equation}
 and its (time) Fourier transform

\begin{equation}
G_{ij}(\omega)=\intop_{-\infty}^{\infty}d\tau\, G_{ij}(\tau)e^{i\omega\tau}\equiv\left\langle \left\langle s_{i}^{+}(\tau);s_{j}^{-}\right\rangle \right\rangle _{\omega}.\label{eq:FTEMGF}
\end{equation}

Using the Poisson brackets for the classical spin variables $\mathbf{s}_{i}$
\cite{kleinert95WS}

\[
\left\{ s_{i}^{\pm},s_{j}^{z}\right\} =\pm i\delta_{ij}s_{i}^{\pm},\quad\left\{ s_{i}^{+},s_{j}^{-}\right\} =-2i\delta_{ij}s_{i}^{z},
\]
 we obtain the equation of motion for $G_{ij}(\tau)$ and thereby
for its Fourier transform $G_{ij}(\omega)$ 
\begin{eqnarray}
-i\omega G_{ij}(\omega) & = & -2i\delta_{ij}\left\langle s_{i}^{z}(0)\right\rangle -i\left(g\mu_{B}H'\right)G_{ij}(\omega)-2iK'\left\langle \mbox{\ensuremath{\left\langle s_{i}^{z}(\tau)s_{i}^{+}(\tau),s_{j}^{-}(0)\right\rangle }}\right\rangle _{\omega}\nonumber \\
 & + & i\sum_{l}J'_{il}\left\langle \mbox{\ensuremath{\left\langle s_{i}^{z}(\tau)s_{l}^{+}(\tau);s_{j}^{-}(0)\right\rangle }}\right\rangle _{\omega}-i\sum_{l}J'_{il}\left\langle \mbox{\ensuremath{\left\langle s_{l}^{z}(\tau)s_{i}^{+}(\tau);s_{j}^{-}(0)\right\rangle }}\right\rangle _{\omega}.\label{eq:GreenClassical}
\end{eqnarray}

Then, in analogy with the quantum-mechanical decoupling of exchange
in Eq. (\ref{eq:DecoupExchTahirKheliCallen}), we propose the following
decoupling scheme

\begin{align}
\left\langle \left\langle s_{i}^{z}(\tau)s_{j}^{+}(\tau);s_{l}^{-}(0)\right\rangle \right\rangle _{\omega} & \simeq\left\langle s_{i}^{z}\right\rangle \left\langle \left\langle s_{j}^{+}(\tau);s_{l}^{-}\right\rangle \right\rangle _{\omega}-\left\langle s_{i}^{z}\right\rangle \frac{\left\langle s_{i}^{+}s_{j}^{-}\right\rangle }{2}\left\langle \left\langle s_{i}^{+}(\tau);s_{l}^{-}\right\rangle \right\rangle _{\omega}.\label{eq:DecClass-exch}
\end{align}
 We will show below that this decoupling scheme leads to the correct
classical limit of the SW dispersion and magnetization.

Similar to Eq. (\ref{eq:DecoupAnisAC}), the decoupling of anisotropy
contributions is obtained from the equation above upon setting $l=i$.
Note that this way the same decoupling scheme applies to both quantum
and classical spins, and to both exchange and anisotropy. As discussed
earlier, for quantum spins this unification of exchange and anisotropy
decoupling schemes is due to the expansion in Eq. (\ref{eq:SzCallenExpansion})
for $S^{z}$. However, on the classical side there is no such expansion.
This is a consequence of the fact that the second identity in Eq.
(\ref{eq:SU(2)Identities}) becomes meaningless owing to $\left[S^{+},S^{-}\right]=0$.

Therefore, applying these two decoupling schemes and passing to the
Fourier space in Eq. (\ref{eq:GreenClassical}) we obtain

\[
G_{\mathbf{k}}(\omega^{\prime})=\frac{2\mathcal{N}m}{\omega^{\prime}-\omega^{\prime}\left(\mathbf{k}\right)}
\]
 with the classical dispersion relation ($\omega_{\mathbf{k}}^{\prime}\equiv\omega^{\prime}\left(\mathbf{k}\right)$)
\begin{align*}
\hbar\omega_{\mathbf{k}}^{\prime} & =g\mu_{B}H'+2K^{\prime}m\left(1-\frac{\left\langle s_{i}^{+}s_{i}^{-}\right\rangle }{2}\right)+m\left[J^{\prime}(0)-J^{\prime}(\mathbf{k})\right]\\
 & +\frac{m}{2\mathcal{N}}\sum_{p}\left[J^{\prime}(\mathbf{p})-J^{\prime}(\mathbf{p}-\mathbf{k})\right]\sum_{i,j}e^{i\mathbf{p}\cdot\mathbf{r}_{ij}}\left\langle s_{j}^{+}s_{i}^{-}\right\rangle .
\end{align*}
 Note that we have used the translational invariance to write $\left\langle s_{i}^{z}\right\rangle =\left\langle s^{z}\right\rangle =m$.

Now if we apply the classical analog of the spectral theorem \cite{zubarev60spu,cavalloetal05review},
\emph{i.e.},

\[
G_{k}(\omega^{\prime}+i\epsilon)-G_{k}(\omega^{\prime}-i\epsilon)=-4i\pi\mathcal{N}m\delta(\omega^{\prime}-\omega_{\mathbf{k}}^{\prime}),
\]
 we obtain

\begin{eqnarray}
\sum_{i,j}e^{i\mathbf{p}\cdot\mathbf{r}_{ij}}\left\langle s_{j}^{+}s_{i}^{-}\right\rangle  & = & \frac{2m}{\beta\hbar\omega_{\mathbf{p}}^{\prime}},\qquad\left\langle s_{i}^{+}s_{i}^{-}\right\rangle =\frac{2m}{\beta\mathcal{N}}\sum_{k}\frac{1}{\hbar\omega_{\mathbf{k}}^{\prime}}.\label{eq:SpSmCorr}
\end{eqnarray}

Inserting these expressions back into $\omega_{\mathbf{k}}^{\prime}$
we obtain the classical analog of the dispersion relation that accounts
for the SW interactions

\begin{eqnarray}
\hbar\omega_{\mathbf{k}}^{\prime} & = & g\mu_{B}H'+2K'm\left[1-\frac{m}{\beta\mathcal{N}}\sum_{\mathbf{p}}\frac{1}{\hbar\omega_{\mathbf{p}}^{\prime}}\right]\label{eq:DispCGF}\\
 &  & +m\left[J'(0)-J^{\prime}(\mathbf{k})\right]+\frac{m^{2}}{\beta\mathcal{N}}\sum_{\mathbf{p}}\left[\frac{J^{\prime}(\mathbf{p})-J^{\prime}(\mathbf{p}-\mathbf{k})}{\hbar\omega_{\mathbf{p}}^{\prime}}\right].\nonumber 
\end{eqnarray}

We stress again that only after solving this transcendental equation,
one obtains the final SW dispersion $\omega_{\mathbf{k}}$. This is,
however, a heavy procedure because $\omega_{\mathbf{k}}$ also enters
the magnetization $m$, which in turn involves $\omega_{\mathbf{k}}$
via $\Omega$, and \emph{vice versa}. At each step one has to compute
three-dimensional sums (or integrals) in Fourier space.

Obviously, this dispersion can also be obtained by taking the classical
limit of the quantum GF result, \emph{i.e.} Eq. (\ref{eq:NLMagnonDispGF}).
Indeed, in the presence of uniaxial anisotropy, the Anderson-Callen
decoupling yields the equation {[}see Ref. \citep{atxitiaetal10prb}
and references therein{]} 
\begin{align*}
\hbar\omega_{\mathbf{k}} & =g\mu_{B}H+2K\left\langle S^{z}\right\rangle \left[1-\frac{1}{2S^{2}}\left(S(S+1)-\left\langle S^{z}S^{z}\right\rangle \right)\right]\\
 & +\left\langle S^{z}\right\rangle \left(J(0)-J(\mathbf{k})\right)+\frac{\left\langle S^{z}\right\rangle ^{2}}{\mathcal{N}S^{2}}\sum_{\mathbf{p}}\left[\frac{J(\mathbf{p})-J(\mathbf{p}-\mathbf{k})}{e^{\beta\hbar\omega_{\mathbf{p}}}-1}\right].
\end{align*}

Then, using the identities (\ref{eq:SU(2)Identities}) and making
the substitutions $\omega_{\mathbf{k}}=\omega_{\mathbf{k}}^{\prime}/S,J_{ij}=J'_{ij}/S^{2},K=K'/S^{2},H=H'/S$,
together with $m=\left\langle S^{z}\right\rangle /S$, we obtain

\begin{eqnarray*}
\hbar\omega'_{\mathbf{k}} & = & g\mu_{B}H'+2K'm\left[1+\frac{m}{2S}-\frac{\left\langle s^{+}s^{-}\right\rangle }{2}\right]+m\left(J'(0)-J'(\mathbf{k})\right)+\frac{m^{2}}{\mathcal{N}}\sum_{\mathbf{p}}\left[\frac{J'(\mathbf{p})-J'(\mathbf{p}-\mathbf{k})}{S\left(e^{\beta\frac{\hbar\omega_{'\mathbf{p}}}{S}}-1\right)}\right].
\end{eqnarray*}

In the classical limit $\left\langle n_{\mathbf{p}}\right\rangle $
in Eq. (\ref{eq:OccupationNumb}) becomes 
\begin{equation}
\left\langle n_{\mathbf{p}}\right\rangle =\frac{1}{e^{\beta\frac{\hbar\omega_{'\mathbf{p}}}{S}}-1}\rightarrow\frac{1}{\beta\frac{\hbar\omega_{'\mathbf{p}}}{S}}\label{eq:MagnonDensityClass}
\end{equation}
 and thereby

\begin{align*}
\left(\hbar\omega'_{\mathbf{k}}\right)_{S\to\infty} & =g\mu_{B}H'+2K'm\left[1-\frac{\left\langle s^{+}s^{-}\right\rangle }{2}\right]+m\left[J'(0)-J'(\mathbf{k})\right]+\frac{m^{2}}{\mathcal{N}}\sum_{\mathbf{p}}\left[\frac{J'(\mathbf{p})-J'(\mathbf{p}-\mathbf{k})}{\beta\hbar\left(\omega'_{\mathbf{p}}\right)_{S\to\infty}}\right].
\end{align*}

Next, upon replacing $\left\langle s_{i}^{+}s_{i}^{-}\right\rangle $
by its expression given in Eq. (\ref{eq:SpSmCorr}) we obtain 
\begin{align}
\left(\hbar\omega'_{\mathbf{k}}\right)_{S\to\infty} & =g\mu_{B}H'+2K'm\left[1-\frac{m}{\beta\mathcal{N}}\sum_{\mathbf{p}}\frac{1}{\left(\hbar\omega'_{\mathbf{p}}\right)_{S\to\infty}}\right]\label{eq:CGFDisp}\\
 & +m\left[J'(0)-J'(\mathbf{k})\right]+\frac{m^{2}}{\mathcal{N}}\sum_{\mathbf{p}}\left[\frac{J'(\mathbf{p})-J'(\mathbf{p}-\mathbf{k})}{\beta\hbar\left(\omega'_{\mathbf{p}}\right)_{S\to\infty}}\right].\nonumber 
\end{align}
 This is the dispersion in Eq. (\ref{eq:DispCGF}), which was obtained
directly from the retarded classical GF (\ref{eq:RetardedCGF}) using
the (classical) decoupling scheme (\ref{eq:DecClass-exch}) for exchange
and its analog for anisotropy. Therefore, starting directly with GFs
for classical spins and using the classical analog of the spectral
theorem leads, as it should, to the same result that is achieved by
proceeding with the GFs for quantum spins and taking the classical
limit at the very end.

Similarly to the quantum case, the dispersion (\ref{eq:CGFDisp})
can be recast in the form 
\[
\left(\hbar\omega'_{\mathbf{k}}\right)_{S\to\infty}=g\mu_{B}H'+2\mathcal{K}^{\prime}m+mJ'(0)Q^{\prime}\left(\alpha,\beta\right)\left(1-\gamma_{\mathbf{k}}\right)
\]
 where we have introduced the classical analogs of the effective anisotropy
(\ref{eq:EffectiveAnisotropy-1}) and the exchange stiffness (\ref{eq:QExpression})
\begin{align*}
\mathcal{K}^{\prime} & \equiv K'\left[1-\frac{m}{\beta\mathcal{N}}\sum_{\mathbf{p}}\frac{1}{\left(\hbar\omega'_{\mathbf{p}}\right)_{S\to\infty}}\right],\quad Q^{\prime}\left(\alpha,\beta\right)\equiv1+\frac{m}{\mathcal{N}}\sum_{\mathbf{p}}\frac{\gamma_{\bm{p}}}{\beta\left(\hbar\omega'_{\mathbf{p}}\right)_{S\to\infty}}.
\end{align*}

Before ending this section we discuss the magnetization. The large-spin
limit, \emph{i.e.} $S\longrightarrow\infty$, yields the classical
limit of the Brillouin function, that is the Langevin function, \emph{i.e.},

\[
\lim_{S\to\infty}B_{S}\left(x\right)=\mathcal{L}\left(x\right)=\coth\left(x\right)-\frac{1}{x}.
\]

On the other hand, this is what one obtains when the quantum spins
are replaced by classical vectors and, in the partition function,
the \emph{trace} operator is replaced by integrals on the spin variables
(or their spherical coordinates). Doing so for independent spins in
a magnetic field $x$ leads to the Langevin function.

Now, in Eq. (\ref{eq:Callen-Brillouin}) setting $m=\left\langle \sigma^{3}\right\rangle /S$
and taking the limit $S\to\infty$ yields the magnetization in the
classical limit, \emph{i.e.} 
\begin{equation}
\lim_{S\to\infty}m=\left\langle s^{z}\right\rangle _{\mathrm{class}}=\mathcal{L}\left(\frac{1}{\rho}\right),\label{eq:MagCGF}
\end{equation}
 with\textcolor{blue}{{} }

\begin{equation}
\rho\equiv\frac{1}{N_{c}}\sum_{\mathrm{\mathbf{k}}}\frac{1}{\beta\hbar\omega_{\mathbf{k}}^{\prime}}\label{eq:rho_class}
\end{equation}
 being the classical density of SW excitations.

We note in passing that it is more straightforward to obtain the classical
limit (\ref{eq:MagCGF}) from Eq. (\ref{eq:Callen-Brillouin}) than
from (\ref{eq:CallenMag}). On the other hand, Eq. (\ref{eq:Callen-Brillouin})
provides a clear connection with MFT. Indeed, as discussed earlier,
this connection can be revealed by noting that all quasi-particle
excitations in MFT are degenerate and thus one can simply drop the
dependence on the wave vector in Eq. (\ref{eq:LAOmega}). However,
this similarity in form should not shadow the fundamental difference,
namely that in pure MFT the magnetization $\left\langle s^{z}\right\rangle _{\mathrm{class}}$
is calculated self-consistently using (in a longitudinal magnetic
field) 
\begin{equation}
\left\langle s^{z}\right\rangle =\mathcal{L}\left[\beta S\left(g\mu_{B}H'^{z}+2K'\left\langle s^{z}\right\rangle +J'\left(0\right)\left\langle s^{z}\right\rangle \right)\right]\label{eq:LangevinMag}
\end{equation}
 while in Eq. (\ref{eq:MagCGF}) one explicitly takes into account
the SW dispersion via $\rho$. This SW density is obtained by the
GF technique using the RPA decoupling for exchange contribution and
the Anderson-Callen decoupling for single-ion anisotropy contribution.
Eq. (\ref{eq:MagCGF}) is also a (self-consistent) transcendental
equation because $\rho$ is a function of the dispersion $\omega_{\mathbf{k}}$.
For a comparison of the corresponding critical temperatures see Appendix
\ref{app:TC}.

\medskip{}

\textbf{Magnetization asymptotes}

The low-temperature asymptote for the magnetization is obtained by
expanding $\rho$ and then the magnetization with respect to $\tau=k_{\mathrm{B}}T/J_{0}$.
Neglecting the terms due to Callen's decoupling for exchange and anisotropy
terms in the dispersion relation (\ref{eq:DispCGF}) we obtain

\[
\hbar\omega_{\mathbf{k}}^{\prime}=g\mu_{B}H'+2K'm+m\left(J'(0)-J^{\prime}(\mathbf{k})\right).
\]
 which may be rewritten as ($J'(0)\equiv J_{0}^{\prime}=zJ^{\prime}$)

\begin{align*}
\hbar\omega_{\mathbf{k}}^{\prime} & =J_{0}^{\prime}\left[h'+\left(1+\kappa\right)m\right]\left(1-\psi\left(m\right)\gamma_{\mathbf{k}}\right)
\end{align*}
 where we have introduced the function 
\begin{equation}
\psi\left(m\right)\equiv\frac{m}{h^{\prime}+\left(1+\kappa\right)m}.\label{eq:Gofm}
\end{equation}
 with 
\[
h^{\prime}\equiv\frac{g\mu_{B}H'}{J_{0}^{\prime}}=\frac{g\mu_{B}H}{J_{0}S}=\frac{h}{S}
\]
 and $2K^{\prime}/J_{0}^{\prime}=2K/J_{0}=\kappa$.

Then, the density $\rho$ in Eq. (\ref{eq:rho_class}) becomes 
\begin{equation}
\rho\equiv\psi\left(m\right)P_{N}\left[\psi\left(m\right)\right]\times\frac{\tau^{\prime}}{m}\label{eq:ClassicalDensityPNG}
\end{equation}
 where $\tau^{\prime}\equiv\tau/S^{2}$ and the function $P_{N}\left[\psi\left(m\right)\right]$
reads 
\[
P_{N}\left[\psi\left(m\right)\right]\equiv\frac{1}{N_{c}}\sum_{\mathrm{\mathbf{k}}}\frac{1}{1-\psi\left(m\right)\gamma_{\mathbf{k}}}\simeq W_{N}-c_{0}\sqrt{1-\psi\left(m\right)}
\]
 which is the analog of (\ref{eq:LatticeGF}) for a finite lattice
of linear size $N$. Asymptotic expressions of the lattice Green function
$P_{N}\left(G\right)$ without the zero mode ($\mathbf{k=0}$), for
free boundary conditions (fbc) and periodic boundary conditions (pbc),
can be found in Refs. \citep{kacgar01physa300,kacgar01epjb}. $W_{N}$
is the well known lattice sum whose large-size (continuous) limit
is the Watson integral $W$, introduced earlier in Eq. (\ref{eq:LatticeGFExpand}).
For a bcc lattice we have 
\[
\left\{ \begin{array}{lllll}
W_{N} & \simeq & W_{\mathrm{bcc}}\left(1-\frac{0.65}{N}\right), &  & \mbox{for fbc}\\
\\
W_{N} & \simeq & W_{\mathrm{bcc}}\left(1-\frac{0.83}{N}\right), &  & \mbox{for pbc}
\end{array}\right.
\]
 and for a sc lattice \cite{kacgar01epjb} 
\[
\left\{ \begin{array}{lllll}
W_{N} & \simeq & W_{sc}+\frac{9\ln\left(1.17N\right)}{2\pi N}, &  & \mbox{for fbc}\\
\\
W_{N} & \simeq & W_{\mathrm{sc}}\left(1-\frac{0.90}{N}\right) &  & \mbox{for pbc}.
\end{array}\right.
\]

$W_{sc}$ and $W_{\mathrm{bcc}}$ are the Watson integrals for the
corresponding lattices and are given after Eq. (\ref{eq:LatticeGFExpand}).

Now, at low temperature the density of SW $\rho$ is small and using
$\mathcal{L}\left(x\right)\simeq1-1/x$ for large $x$ in Eq. (\ref{eq:MagCGF})
we obtain the asymptote for the magnetization (up to $2^{\mathrm{nd}}$
order in $\tau$), upon expanding $\rho$ around $m\simeq1$, 
\begin{equation}
m^{\mathrm{CGF}}\simeq1-\rho\simeq1-\psi\left(1\right)P_{N}\left[\psi\left(1\right)\right]\times\tau^{\prime}-\left(\psi\left(1\right)P_{N}\left[\psi\left(1\right)\right]\right)^{2}\left(\tau^{\prime}\right)^{2}.\label{eq:CGFLowTAsympt}
\end{equation}
 Note that $\psi\left(1\right)$ is a function of the applied field,
since according to Eq. (\ref{eq:Gofm}), $\psi\left(1\right)=1/\left(h^{\prime}+1+\kappa\right)$.
To first order in $\tau^{\prime}$ Eq. (\ref{eq:CGFLowTAsympt}) obviously
recovers the low-temperature linear decay of the magnetization which
is typical of the classical Dirac-Heisenberg models. At very low temperature
we can neglect the second-order terms and expand with respect to the
field $h^{\prime}$ leading to

\[
m^{CGF}\simeq1-\frac{W_{N}}{1+\kappa}\tau^{\prime}+\frac{c_{0}}{1+\kappa}\sqrt{\frac{h'+\kappa}{1+\kappa}}\tau^{\prime}.
\]

This is also the SW theory result obtained in Ref. \cite{kacgar01epjb},
second line of Eq. (65), in the absence of anisotropy. We remark in
passing that in this reference SW theory was extended to account for
finite-size effects in fine magnetic particles. One of the consequences
of these effects is that there appears a critical field $H_{V}\sim T/\mathcal{N}$,
that corresponds to the suppression of the global rotation of the
particle's net magnetic moment, below which the magnetization is quadratic
in the applied magnetic field. In the present work, the system size
$\mathcal{N}$ is big enough so that $H_{V}$ vanishes and the quadratic
behavior of the magnetization is suppressed. A more thorough comparison
of the present work with that of Ref. \cite{kacgar01epjb} will be
addressed in a future work.

Near the critical temperature and in the absence of the applied field
we have 
\begin{equation}
\rho=\lambda^{\prime}P\left(\lambda^{\prime}\right)\frac{\tau^{\prime}}{m}.\label{eq:rhoTC}
\end{equation}
 We note that we have replaced the finite-sum lattice Green function
$P_{N}\left(\lambda^{\prime}\right)$ by its continuum limit defined
in (\ref{eq:LatticeGF}) as this is appropriate in the present high
temperature regime. In the absence of the magnetic field and neglecting
Callen's decoupling for anisotropy and exchange, \emph{i.e.} for $h'=0$,
we have $\psi(m)=1/\left(1+\kappa\right)$. Likewise, $\eta$ in Eq.
(\ref{eq:NearTcDispParams}) is simply replaced by one and thereby
$\psi$ equals the parameter $\lambda$ but here with the {}``primed''
parameters, \emph{i.e.} $\lambda\rightarrow\lambda^{\prime}=1/\left(1+\kappa\right)$.

Then, since the magnetization $m$ is small the density of SW $\rho$
is large. Hence, using $\mathcal{L}\left(x\right)\simeq x/3-x^{3}/45$,
for small $x$, and solving for $m^{\mathrm{CGF}}$ we obtain the
asymptotic expression 
\begin{align}
m^{\mathrm{CGF}} & \simeq\sqrt{15}\lambda^{\prime}P\left(\lambda^{\prime}\right)\tau^{\prime}\sqrt{1-3\lambda^{\prime}P\left(\lambda^{\prime}\right)\tau^{\prime}}.\label{eq:CGFHighTAsympt}
\end{align}

As in the quantum case, it is possible to take into account the magnon-magnon
interactions from the last term in Eq. (\ref{eq:DispCGF}). The magnetization
is then given by the following expression

\begin{align}
m^{\mathrm{CGF}} & \simeq\frac{\sqrt{15}\Lambda P\left(\Lambda\right)\tau^{\prime}}{Q^{\prime}(\alpha,\tau^{\prime})}\sqrt{1-\frac{3\Lambda P\left(\Lambda\right)\tau^{\prime}}{Q^{\prime}(\alpha,\tau^{\prime})}},\label{eq:CGFHightTAsymptQ}
\end{align}
 with $\alpha=0,m^{2},m^{4},m\left(m+m^{3}\right)$ for the RPA, Callen,
Copeland-Gersch or Swendsen decoupling, respectively. Now we can see
that this can be recovered, as it should, as the classical limit of
the asymptote (\ref{eq:MagCGFNearTCPhi}) obtained for quantum spins.
Indeed replacing the various parameters by their classical counterparts
(\emph{e.g.} $J_{0}$ by $J_{0}^{\prime}=J_{0}S^{2}$, $\tau^{\prime}=\tau/S^{2}$,
etc.) and dividing $\bar{\sigma}$ by $S$ we obtain

\[
\frac{\bar{\sigma}^{\textrm{QGF}}}{S}\simeq\frac{\sqrt{15}\Lambda P(\Lambda)\tau^{\prime}}{Q(\alpha,\tau^{\prime})\sqrt{1+\frac{1}{S^{2}}\left(1+\frac{5P(\Lambda)-3}{4}\right)}}\sqrt{1-\frac{3\Lambda P(\Lambda)\tau^{\prime}}{\left(1+\frac{1}{S^{2}}\right)Q(\alpha,\tau^{\prime})}}.
\]

This readily yields the asymptote in Eq. (\ref{eq:CGFHighTAsympt})
upon taking the limit $S\rightarrow\infty$, here and in Eq. (\ref{eq:LQ}),
and writing $m^{\mathrm{CGF}}=\lim_{S\rightarrow\infty}\left(\bar{\sigma}^{\textrm{QGF}}/S\right)$.

We note that while submitting the present work we became aware of
a recent work \cite{campanaetal11physa} where the classical GF method
is developed along the procedure employed by Callen for quantum spins
based on the generalized GF in Eq. (\ref{eq:CallenGF}). The results
obtained by the authors for the dispersion and magnetization are quite
similar to ours. We stress, however, that the classical GF method
we develop here is more straightforward as it avoids the difficult
algebra involved in the calculation of the GF (\ref{eq:CallenGF}),
which was introduced for dealing with arbitrary quantum spin $S$
\cite{akhbarpel68}. Moreover, our approach is based on a unified
decoupling scheme for both exchange and anisotropy contributions and
establishes a clear connection with the quantum-mechanical Callen's
decoupling. In fact, the work in Ref. \cite{campanaetal11physa} about
Callen's method together with the present approach provide a complete
picture of the GF technique for classical spins.

\subsection{\label{sub:CSD}Classical spectral density method}

In this section we summarize the basic ideas and formulas of the classical
analog of the spectral-density method, the so-called classical spectral
density method (CSD). One of the objectives of this method is to provide
systematic and non trivial approximations in classical statistical
physics when applied to classical spin systems. To the best of our
knowledge, this was initially formulated in \citep{auriaetal81pla}
and later developed and applied by several authors {[}see Refs. \citep{campanaetal84prb,cavalloetal05review}
and references therein{]}. This approach is then compared to the classical
GF technique developed in the previous section. In Ref. \cite{campanaetal11physa},
the CSD method was compared to the classical analog of Callen's method.

Here the spin $\mathbf{s}_{i}$ is a classical vector and the magnetization
is defined by $\bar{m}=\left\langle s^{z}\right\rangle $. One then
defines the classical spectral density $\Lambda_{\mathbf{k}}\left(\omega\right)$
of the time-dependent spin correlations. Then, the calculations proceed
by assuming a given form (\emph{e.g.} a Gaussian or a Lorentzian)
for $\Lambda_{\mathbf{k}}\left(\omega\right)$ involving some parameters
\citep{auriaetal81pla}. The latter are obtained by solving a hierarchy
of (moment) equations which are in turn obtained from a chain of equations
for Green's functions of all orders \citep{auriaetal81pla}. For the
Hamiltonian $\mathcal{H}$ in Eq. (\ref{eq:Hamiltonian}) one obtains
the following dispersion relation \cite{auriaetal81pla,atxitiaetal10prb}
\begin{equation}
\hbar\omega'_{\mathbf{k}}=h'+\frac{1}{\mathcal{N}^{2}}\sum_{\mathbf{q}}\left[\left(-2k'+J'_{\mathbf{q}}-J'_{\mathbf{k}-\mathbf{q}}\right)\left\langle s_{\mathbf{q}}^{+}s_{-\mathbf{q}}^{-}\right\rangle +2\left(2k'+J'_{\mathbf{q}}-J'_{\mathbf{k}-\mathbf{q}}\right)\left\langle s_{\mathbf{q}}^{z}s_{-\mathbf{q}}^{z}\right\rangle \right].\label{eq:DispCSD}
\end{equation}
 This involves the two correlation functions $\left\langle s_{\mathbf{k}}^{z}s_{-\mathbf{k}}^{z}\right\rangle $
and $\left\langle s_{\mathbf{k}}^{+}s_{-\mathbf{k}}^{-}\right\rangle $
which have to be dealt with in order to proceed any further. $\left\langle s_{\mathbf{k}}^{+}s_{-\mathbf{k}}^{-}\right\rangle $
is easily obtained as \cite{atxitiaetal10prb} 
\[
\left\langle s_{\mathbf{k}}^{+}s_{-\mathbf{k}}^{-}\right\rangle =\frac{2\mathcal{N}m}{\beta\hbar\omega'_{\mathbf{k}}}.
\]

The second approximation made in CSD - the first being the form chosen
for the spectral density $\Lambda_{\mathbf{k}}\left(\omega\right)$-
concerns the unavoidable decoupling scheme that is required for the
calculation of the longitudinal correlation function $\left\langle s_{\mathbf{k}}^{z}s_{-\mathbf{k}}^{z}\right\rangle $.
Here we stress that, as is seen from Eq. (\ref{eq:DispCSD}), this
contribution stems from the exchange as well as the anisotropy contribution.
However, as discussed earlier, the decoupling that should be applied
to the one or to the other contribution is rather different from the
physical point of view since this depends on whether this contribution
is a local or a bi-local term. Hence, let us summarize the results
of our developments concerning this issue, which is extremely important
as the soundness of the results is strongly dependent on its outcome.

In fact, to obtain a decoupling for $\left\langle s_{\mathbf{k}}^{z}s_{-\mathbf{k}}^{z}\right\rangle $
that stems from the exchange contribution, we may start from Eq. (\ref{eq:GreenClassical})
use the decoupling (\ref{eq:DecClass-exch}), and then Fourier transform
the result. These developments are carried out in Appendix \ref{app:CSDMExchAnisDecoup}
and their outcome is the following exchange decoupling scheme 
\begin{align}
\sum_{\mathbf{q}}\left(J_{\mathbf{\mathbf{q}}}^{\prime}-J_{\mathbf{\mathbf{k}}-\mathbf{\mathbf{q}}}^{\prime}\right)\left\langle s_{\mathbf{q}}^{z}s_{-\mathbf{q}}^{z}\right\rangle \simeq\sum_{\mathbf{q}}\left(J_{\mathbf{\mathbf{q}}}^{\prime}-J_{\mathbf{\mathbf{k}}-\mathbf{\mathbf{q}}}^{\prime}\right)\left[\left\langle s_{\mathbf{q}}^{z}\right\rangle \left\langle s_{-\mathbf{q}}^{z}\right\rangle -\frac{1}{2}\left(1-m^{2}\right)\left\langle s_{\mathbf{q}}^{+}s_{-\mathbf{q}}^{-}\right\rangle \right].\label{eq:DecouplingExchangeCSD}
\end{align}
 These calculations provide a clear derivation of the exchange decoupling
used in the literature, see \emph{e.g.} Ref. \onlinecite{cavalloetal05review}.
Note, however, that the decoupling of the longitudinal correlation
function (\ref{eq:DecouplingExchangeCSD}) is only valid under the
sum over the wave vector $\mathbf{q}$.

For the anisotropy contribution one may start from the decoupling
scheme\emph{ }proposed in Eq. (\ref{eq:DecClass-exch}) with $j=i$
and decouple the high-order contributions, \emph{i.e.}, 
\begin{align*}
\left\langle \left\langle s_{-\mathbf{q}}^{z}(\tau)s_{-\mathbf{k}}^{+}(\tau);s_{\mathbf{q}+\mathbf{k}}^{-}(0)\right\rangle \right\rangle _{\omega} & =\left\langle s_{-\mathbf{q}}^{z}\right\rangle \left(1-\frac{\left\langle s^{+}s^{-}\right\rangle }{2}\right)\left\langle \left\langle s_{-\mathbf{k}}^{+}(\tau);s_{\mathbf{q}+\mathbf{k}}^{-}\right\rangle \right\rangle _{\omega}.
\end{align*}

In Ref. \citep{campanaetal84prb} the higher-order spectral density
was reduced to a simple form that leads to the correct results in
the low- and high-temperature limits. This yields the following equation
\begin{equation}
m^{2}\left(1-\frac{\left\langle s^{+}s^{-}\right\rangle }{2}\right)=1-\frac{3}{2}\left\langle s^{+}s^{-}\right\rangle .\label{eq:eqcampana}
\end{equation}

In turn this renders the following expression for the magnetization

\begin{equation}
m\simeq\sqrt{\frac{1-3m\rho}{1-m\rho}}.\label{eq:MagCSD}
\end{equation}
 with $\rho$ being the spectral density defined in Eq. (\ref{eq:rho_class}).

Here a remark is in order concerning CSD as compared to CGF. For a
longitudinal magnetic field, Callen's expression (\ref{eq:CallenMag})
or (\ref{eq:Callen-Brillouin}) for the magnetization is exact, whereas
expression (\ref{eq:MagCSD}) rendered by CSD is an approximation.
Hence, in addition to the common approximation related with the decoupling
scheme and which yields the SW dispersion, CSD introduces an additional
approximation for the magnetization itself.

The classical analog of (\ref{eq:SU(2)Identities}) is obtained by
using the condition $\left|\mathbf{s}\right|=1$ and the fact that
for classical spins we have $s^{+}s^{-}=s^{-}s^{+}$. That is

\[
\left\langle s^{z}s^{z}\right\rangle =1-\left\langle s^{+}s^{-}\right\rangle .
\]

Consequently, Eq. (\ref{eq:eqcampana}) can be rewritten as

\[
m^{2}\left(1-\frac{\left\langle s^{+}s^{-}\right\rangle }{2}\right)=\left\langle s^{z}s^{z}\right\rangle -\frac{1}{2}\left\langle s^{+}s^{-}\right\rangle 
\]
 leading to the following decoupling for the anisotropy contribution

\begin{equation}
\left\langle s_{\mathbf{\mathbf{q}}}^{z}s_{-\mathbf{\mathbf{q}}}^{z}\right\rangle \approx m^{2}+\frac{1}{2}\left(1-m^{2}\right)\left\langle s_{\mathbf{\mathbf{q}}}^{+}s_{-\mathbf{\mathbf{q}}}^{-}\right\rangle .\label{eq:DecouplingAnisoCSD}
\end{equation}

As stressed earlier, we see that for the same longitudinal correlation
$\left\langle s_{\mathbf{\mathbf{q}}}^{z}s_{-\mathbf{\mathbf{q}}}^{z}\right\rangle $
we have a different decoupling scheme according to whether this results
from exchange or anisotropy. Notice the difference in sign between
Eq. (\ref{eq:DecouplingExchangeCSD}) and Eq. (\ref{eq:DecouplingAnisoCSD}).

Applying the decoupling (\ref{eq:DecouplingExchangeCSD}) for the
exchange and (\ref{eq:DecouplingAnisoCSD}) for the anisotropy contributions
to Eq. (\ref{eq:DispCSD}) we obtain the expression for the dispersion
that coincides with the classical limit in Eq. (\ref{eq:DispCGF}).
Summarizing, we see that only upon clearly identifying the origin
(exchange or anisotropy) of the correlation function and applying
the right decoupling scheme does one show that the CSD method renders
the same results as the CGF technique. Next, we deal with the magnetization.

\medskip{}

\textbf{Low temperature asymptote}

Expanding Eq. (\ref{eq:MagCSD}) with respect to $\rho$ which is
small here, we obtain 
\[
m\simeq1-\rho m-\frac{3}{2}\rho^{2}m^{2}
\]
 and then using the expression (\ref{eq:ClassicalDensityPNG}) for
$\rho$ we get

\begin{align*}
m & \simeq1-\psi\left(1\right)P_{N}\left[\psi\left(1\right)\right]\tau-\frac{3}{2}\left(\psi\left(1\right)P_{N}\left[\psi\left(1\right)\right]\right)^{2}\tau^{2}.
\end{align*}

Upon setting $m\sim1$ in the right-hand side we see that this expression
and the corresponding CGF asymptote (\ref{eq:CGFLowTAsympt}) differ
only at the second order in $\tau^{\prime}$ by a factor of $3/2$.
In the case $h^{\prime}=0$ we obtain

\begin{align*}
m^{\mathrm{CSD}} & \simeq1-\lambda^{\prime}P\left(\lambda^{\prime}\right)\tau^{\prime}-\frac{3}{2}\left(\lambda^{\prime}P\left(\lambda^{\prime}\right)\tau^{\prime}\right)^{2}.
\end{align*}

\medskip{}

\textbf{Near-critical temperature asymptote ($h^{\prime}=0$)}

Starting again from Eq. (\ref{eq:MagCSD}) in the absence of magnetic
field and using the expression (\ref{eq:rhoTC} ) for $\rho$, we
obtain

\begin{equation}
m^{\mathrm{CSD}}\simeq\sqrt{\frac{1}{1-\lambda^{\prime}P\left(\lambda^{\prime}\right)\tau^{\prime}}}\sqrt{1-3\lambda^{\prime}P\left(\lambda^{\prime}\right)\tau^{\prime}}.\label{eq:CSDHighTAsympt}
\end{equation}

Similarly to the CGF method, if we take into account the magnon-magnon
interactions by introducing the parameter $\alpha$, the magnetization
becomes

\[
m^{\mathrm{CSD}}\simeq\sqrt{\frac{1}{1-\frac{\Lambda P\left(\Lambda\right)\tau^{\prime}}{Q^{\prime}(\alpha,\tau^{\prime})}}}\sqrt{1-\frac{3\Lambda P\left(\Lambda\right)\tau^{\prime}}{Q^{\prime}(\alpha,\tau^{\prime})}}.
\]

For comparison, we give the following relation between the CGF and
CSD high-temperature asymptotes

\begin{align*}
m^{\mathrm{CGF}} & \simeq\frac{\sqrt{15}\Lambda P\left(\Lambda\right)\tau^{\prime}}{Q^{\prime}(\alpha,\tau^{\prime})}\sqrt{1-\frac{\Lambda P\left(\Lambda\right)\tau^{\prime}}{Q^{\prime}(\alpha,\tau^{\prime})}}\times m^{\mathrm{CSD}}.
\end{align*}

\subsection{Numerical methods}

One of the numerical techniques used here is based on the Langevin
dynamics simulations of thermally excited spin waves \citep{chubykaloPRB02,atxitiaetal10prb}
in the classical case. The method is based on the numerical integration
of the stochastic LLE

\begin{equation}
\frac{d\bm{\mathbf{s}}_{i}}{dt}=-\frac{\gamma}{\mu}\Big(\bm{\mathbf{s}}_{i}\times\bm{\mathbf{H}}_{i}^{{\textrm{eff}}}+\lambda\bm{\mathbf{s}}_{i}\times\left[\bm{\mathbf{s}}_{i}\times\bm{\mathbf{H}}_{i}^{{\textrm{eff}}}\right]\Big)\label{eq:llg}
\end{equation}
 where $\bm{s}$ is the classical localized spin corresponding to
a localized magnetic moment with modulus $\mu$. $\lambda$ and $\gamma$
are the Gilbert damping parameter and the gyromagnetic ratio respectively.
The effective field, $\bm{\mathbf{H}}_{i}^{\mathrm{eff}}$, is then
given by: 
\begin{equation}
\bm{\mathbf{H}}_{i}^{\textrm{eff}}=\bm{\zeta}_{i}(t)-\frac{1}{\mu}\frac{\partial\mathcal{H}_{i}}{\partial\bm{\mathbf{s}}_{i}}.\label{eq:deriv_ham}
\end{equation}
 Here $\bm{\zeta}_{i}(t)$ is the stochastic term that describes the
coupling to the external heat bath. Thermal fluctuations are included
as a white noise term (uncorrelated in time) which is added into the
effective field. The thermal fields are calculated by generating Gaussian
random numbers and multiplying by the strength of the noise process.
The correlators of different components of this field are given by

\begin{equation}
\left\langle \bm{\zeta}_{i,\alpha}(t)\bm{\zeta}_{j,\beta}(t')\right\rangle =\frac{2\lambda k_{\text{B}}T}{\mu\gamma}\delta_{ij}\delta_{\alpha\beta}\delta(t-t')\label{eq:white_noise_correlator}
\end{equation}
 where $\alpha,\beta$ refer to the Cartesian components and $T$
is the temperature of the heat bath to which the spin is coupled.

Using this technique, we simulate a generic three dimensional ferromagnet
with a Heisenberg Hamiltonian as in Eq.\ (\ref{eq:Hamiltonian})
with an external applied field $H$ parallel to the $z-$axis. The
correlated magnetization fluctuations introduced by the random Langevin
field are dealt with by Fourier analysis, both in space and time.
More precisely, we transform the magnetization fluctuations $\mathbf{\widetilde{s}}\left(\mathbf{r},t\right)=\left(s_{x}\left(\mathbf{r},t\right),s_{y}\left(\mathbf{r},t\right)\right)$
around the equilibrium direction along the axis $z$ via a Discrete
Fourier Transform $\mathcal{DFT}$, 
\begin{equation}
\widetilde{\mathbf{s}}(\mathbf{k},\omega_{n})=\mathcal{DFT}\left(\mathbf{\widetilde{s}}\left(\mathbf{r},t_{n}\right)\right)\label{eq:FTMag}
\end{equation}
 where $\{t_{n}\}$ are discrete time instants and the wave vector
for a finite box-shaped ferromagnet with periodic boundary conditions
takes the form \citep{kacgar01physa300,kacgar01physa291} $ak_{\alpha}=2\pi n_{\alpha}/N_{\alpha}$
with $n_{\alpha}=0,1,\ldots,N_{\alpha}-1$; $\alpha=x,y,z$. Then
we compute the power spectrum density defined by $F(\mathbf{k},w)=|\widetilde{\mathbf{s}}(\mathbf{k},\omega)|^{2}$.

The second numerical method used in this work is the classical Monte
Carlo simulation technique using the standard Metropolis algorithm,
see \emph{e.g.} Refs. \citep{binher92, kacgar01epjb}. The results
of this method are used as a benchmark for those rendered by the (semi-)analytical
methods of QGF/CGF and CSD with various decoupling schemes. For equilibrium
properties it is well known that MC and LLE render similar results,
with the difference that the former method is computationally faster
at high temperatures whereas at low temperature LLE is faster. At
the same time, we should note that the MC techniques do not include
proper magnetization dynamics and thus are not suitable for the calculations
of the spin wave spectrum but certainly are for the magnetization.

\section{Results and further comparison between different methods.}

In this section we present a sample of the results for the SW spectrum
and magnetization as a function of temperature and magnetic field,
taking account of magnon-magnon interactions within various decoupling
schemes. The second objective here is to compare the latter and assess
their validity. We also evaluate the temperature-dependent exchange
stiffness and provide (in Appendix \ref{app:TC}) analytical expressions
for the Curie temperature within the decoupling schemes considered.

\subsection{\label{sec:Magnetization-curves}Temperature-dependent magnetization
within different decoupling schemes}

\begin{figure}[H]
\begin{centering}
\includegraphics[width=10cm,height=7cm]{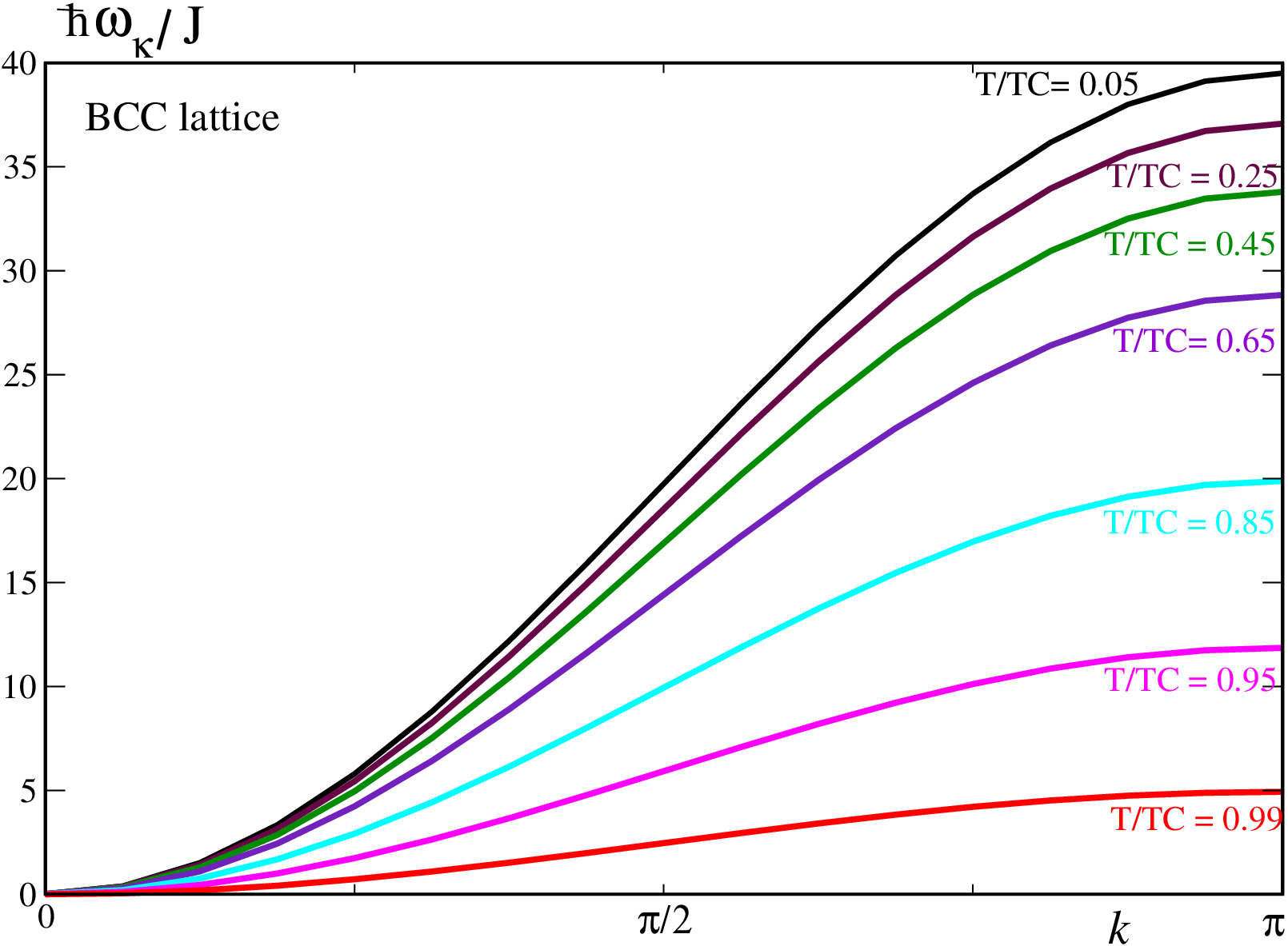} 
\par\end{centering}

\caption{\label{fig:Dispersion}Dispersion relations, obtained by the CGF method,
with wave vector (0,0,k) and without magnetic field.}
\end{figure}

First, as an illustration of the temperature dependence of the SW
spectrum, we plot in Fig. \ref{fig:Dispersion} the dispersion as
a function of the wave vector $\mathbf{k}$ along the $z$ axis, for
different temperatures. 

It can be seen that $\omega_{\mathbf{k}}$, which includes magnon-magnon
interactions, is strongly dependent on temperature. At temperatures
near the critical value, the SW softening is clearly seen. A favorable
comparison of these curves obtained by the CSD method with those rendered
by the numerical LLE method was presented in Ref. \citep{atxitiaetal10prb}.

Now, we present the magnetization curves, as a function of temperature
and applied field, computed with the different methods for the bcc
lattice and iron parameters (per atom) $J=1.44\:10^{-21}{\rm J}$
and $K=5.4\:10^{-24}{\rm J}$. 

\begin{figure}[H]

\begin{centering}
\includegraphics[width=12cm,height=8cm]{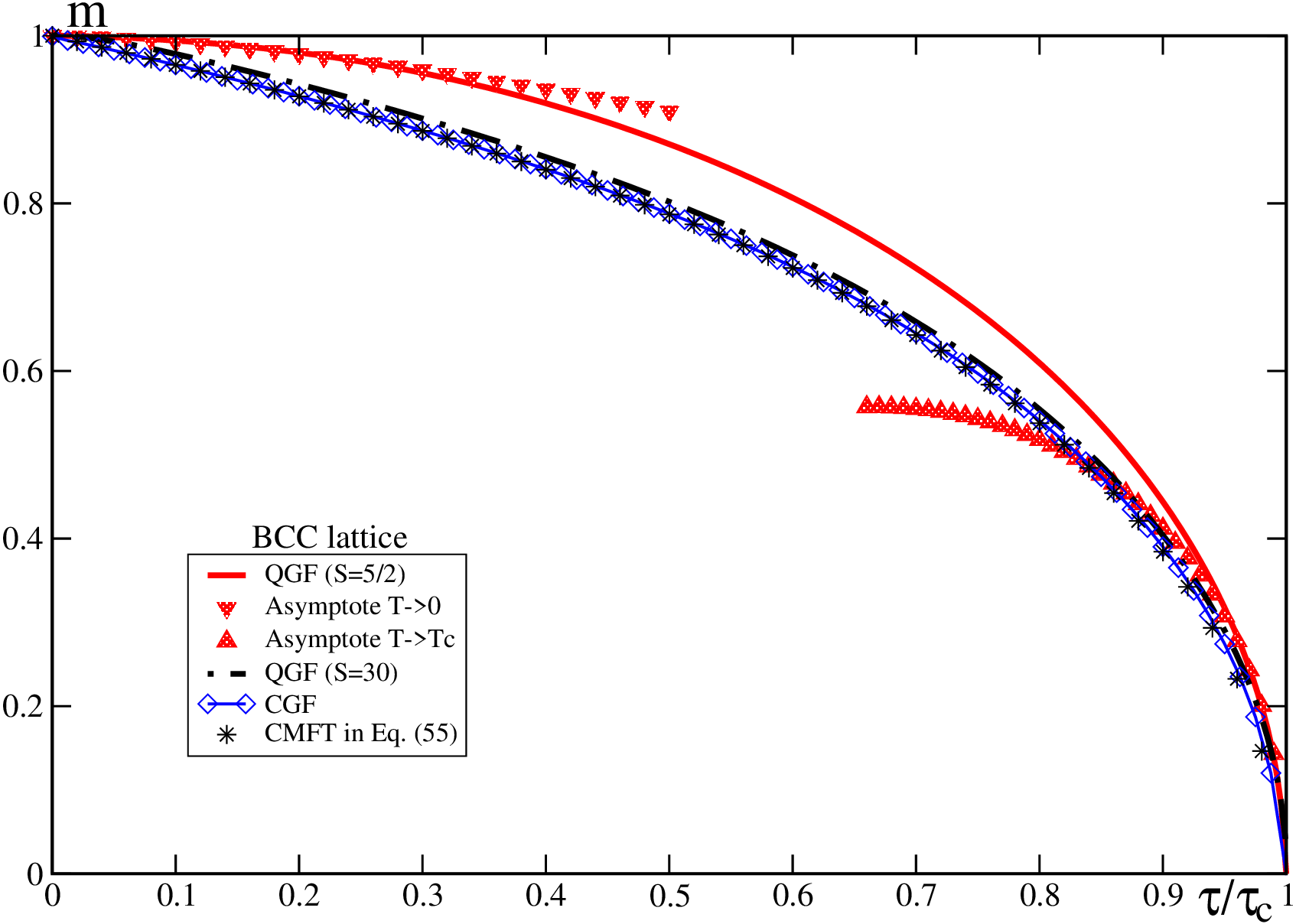} 
\par\end{centering}

\caption{\label{fig:m(T)QGF2CGF}Magnetization against reduced temperature.
Comparison of i) QGF for $S=5/2,30$, ii) CGF, and iii) classical
MFT. We also show the low-temperature asymptote (\ref{eq:LowTMag})
and the asymptote near the critical temperature (\ref{eq:MagQGFNearTc}).
The critical temperature $\tau_{c}$ is that of the method used for
obtaining the corresponding magnetization curve and $\tau\equiv k_{B}T/J_{0}$.}
\end{figure}

In Fig. \ref{fig:m(T)QGF2CGF}, we plot the magnetization $m=\left\langle S^{z}\right\rangle /S$
as a function of (reduced) temperature $\tau=k_{\mathrm{B}}T/J_{0}$
in zero magnetic field, as obtained from i) QGF with two values of
the nominal spin $S=5/2,30$, from ii) CGF, and from iii) classical
MFT (CMFT), \emph{i.e. }Eq. (\ref{eq:LangevinMag}). We see that as
$S$ increases the magnetization curve $m$ tends to that rendered
by CGF and CMFT. In particular, at low temperature we do see the evolution
from the $m\sim T^{3/2}$ Bloch law to the linear law $m\sim T$,
as is typical of the classical Dirac-Heisenberg model. It is interesting
how the CMFT result agrees with that of CGS when $m$ is plotted against
the reduced temperature. The low-temperature asymptote (\ref{eq:LowTMag})
shows a good agreement with the QGF curve for $T\lesssim T_{\mathrm{C}}/4$.
Similarly, the asymptote in the critical region (\ref{eq:MagQGFNearTc})
also reproduces correctly the QGF curve.

\begin{figure}[H]

\begin{centering}
\includegraphics[width=10cm,height=8cm]{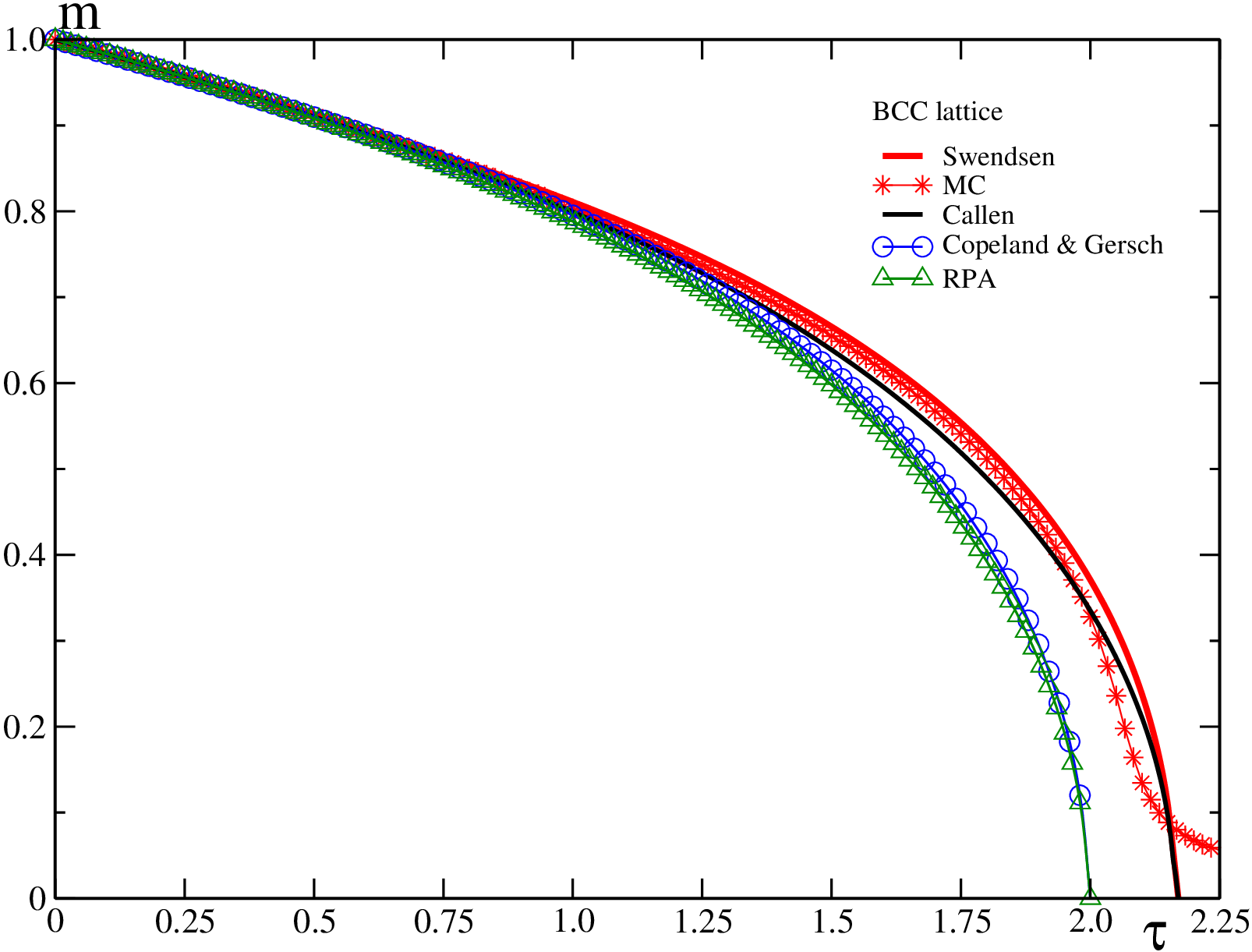} 
\par\end{centering}

\caption{\label{fig:CGF-decoupling}Magnetization curves rendered by different
decoupling schemes. The methods are compared to Monte Carlo.}
\end{figure}

In Fig. \ref{fig:CGF-decoupling}, we compare the magnetization curves
rendered by CGF {[}see Eq. (\ref{eq:MagCGF}){]} for various decoupling
schemes, with MC as a benchmark. Here, we prefer to plot the magnetization
against the absolute temperature $\tau$ so as to see how different
are the critical values of temperature rendered by the various decoupling
methods.

It is seen that the decoupling schemes of Callen and Swendsen agree
quite well with MC. On the other hand, Copeland $\&$ Gersch (CG)
and RPA decoupling schemes render nearly the same curve $m\left(\tau\right)$
that goes below the previous curves at high temperature. This is simply
due to the fact that decoupling schemes with terms of high powers
of $m$, \emph{e.g. }$3$ in the CG decoupling and in the second term
in Swendsen's decoupling {[}see Eq. (\ref{eq:SwendsenDecoup}){]},
lead to a negligible contribution at temperatures nearing the critical
value. On the contrary, contributions that are linear in $m$ in the
decoupling schemes, such as Callen's and Swendsen's, do improve the
magnetization curve at all temperatures.

\begin{figure}[H]

\begin{centering}
\includegraphics[width=10cm,height=7cm]{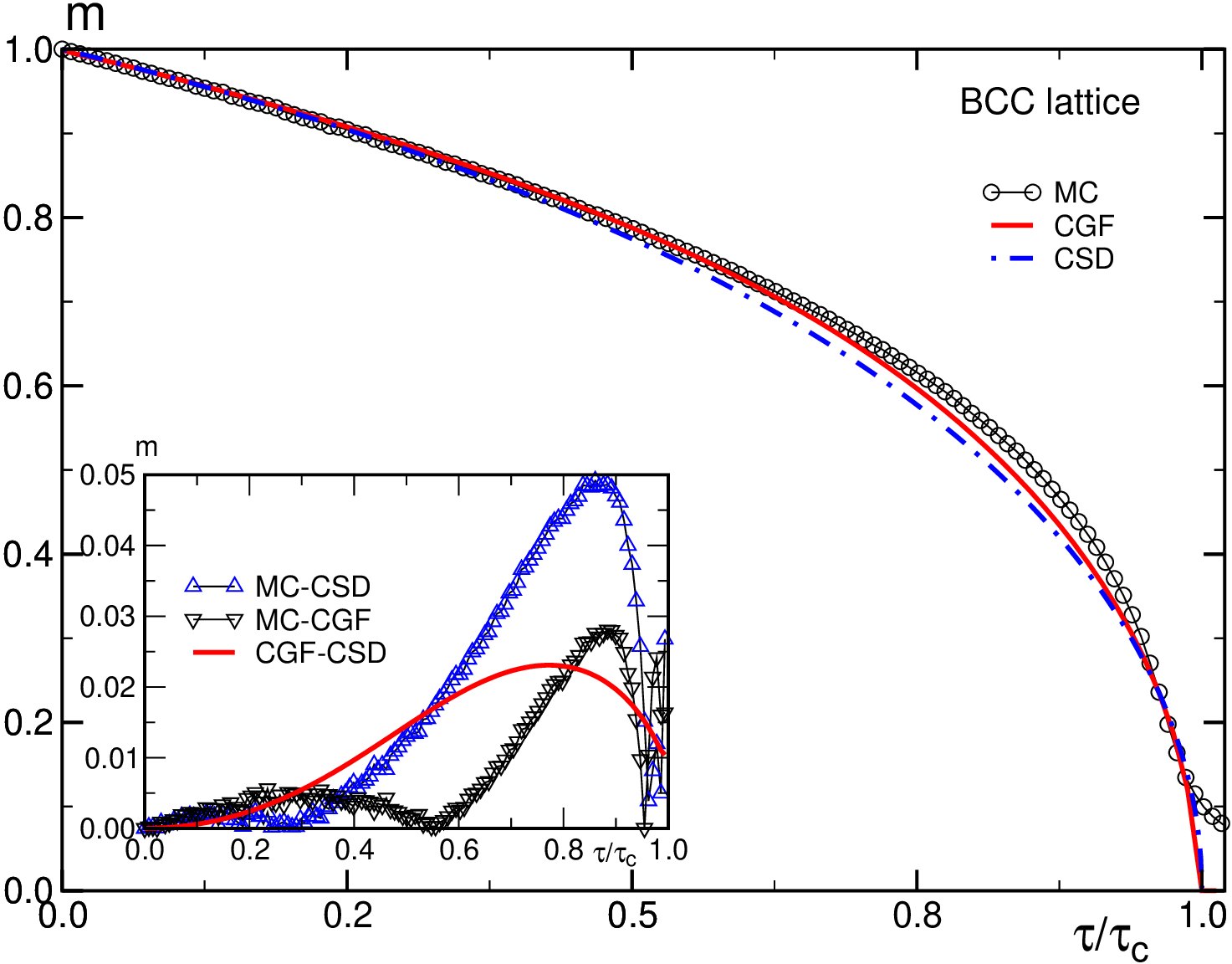} 
\par\end{centering}

\caption{\label{fig:CGF-CSD-MC}Magnetization obtained by CGF and CSD within
RPA. The two methods are compared to Monte Carlo. Inset: difference
between CGF, CSD (with RPA decoupling) and MC.}
\end{figure}

In Fig. \ref{fig:CGF-CSD-MC}, we compare the magnetization rendered
by i) CGF and its Langevin function in Eq. (\ref{eq:MagCGF}) and
ii) CSD given by Eq. (\ref{eq:MagCSD}), within RPA, and the two results
are compared to MC. Globally, CGF renders a magnetization curve that
keeps closer to MC than CSD method, which does so only at low temperature
and near the critical temperature.

In the inset we plot the three differences between the CSD, CGF, and
MC. It is seen that large deviations occur for $T/T_{C}\gtrsim0.4$.

\begin{figure}[H]

\begin{centering}
\includegraphics[width=10cm,height=7cm]{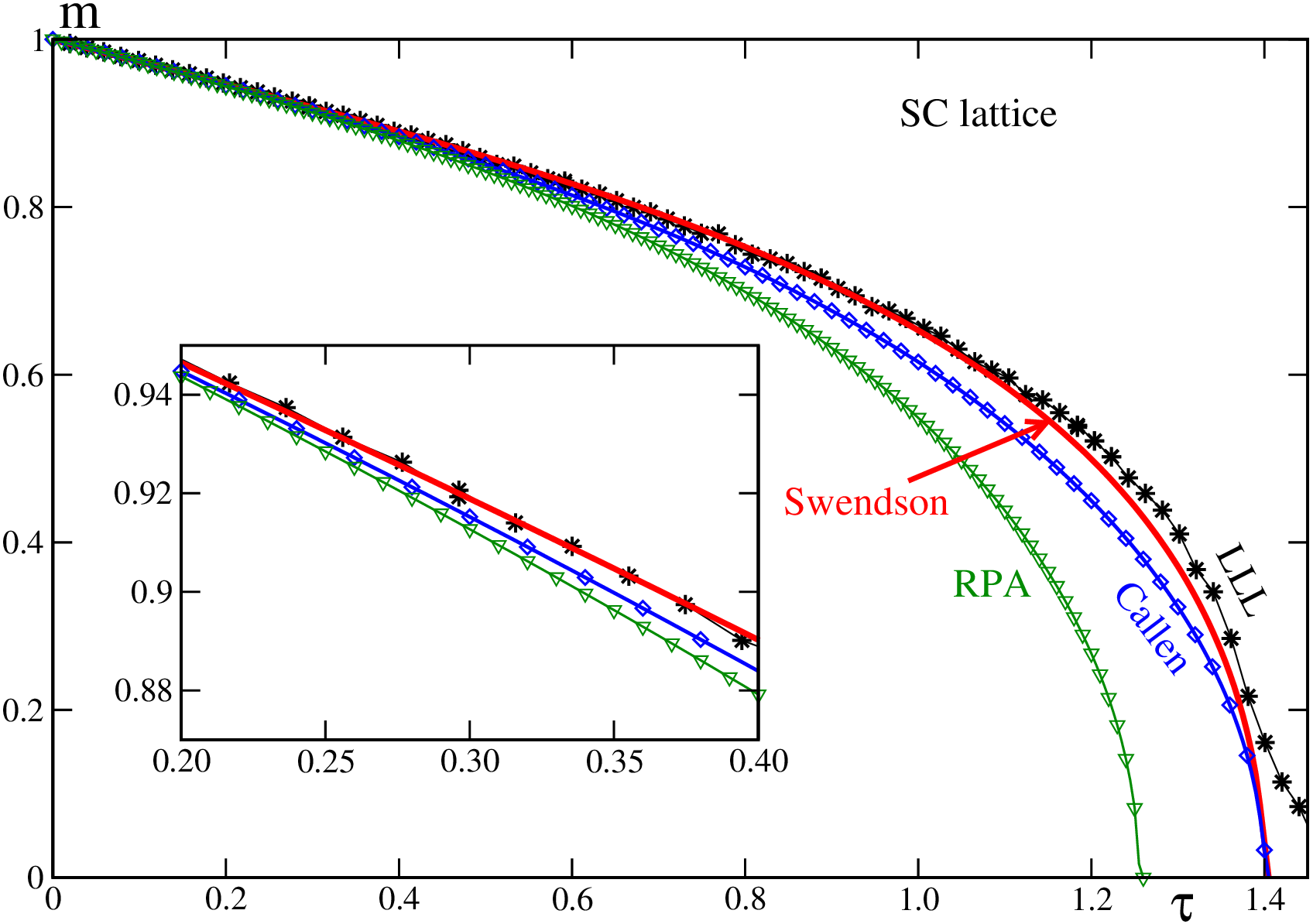} 
\par\end{centering}

\caption{\label{fig:m(T)QGF2CGF_SC}Comparison between the GF methods, the
CSD method with RPA decoupling, the LLE approach and the classical
Green's function for an SC lattice structure.}
\end{figure}

In Fig. \ref{fig:m(T)QGF2CGF_SC} we compare, for the simpler case
of an sc lattice, the classical Green's function method with three
decoupling schemes, with the numerical LLE method. It is seen that
LLE compares quite well with CGF using the Swendson decoupling in
almost the whole range of temperature. Note however, that in the numerical
LLE method the finite-size effects are clearly seen in the critical
temperature region, as is also the case with MC, while the analytical
methods do not ignore such effects for they implicitly consider an
infinite lattice.

\begin{figure}[H]

\begin{centering}
\includegraphics[width=8cm,height=6cm]{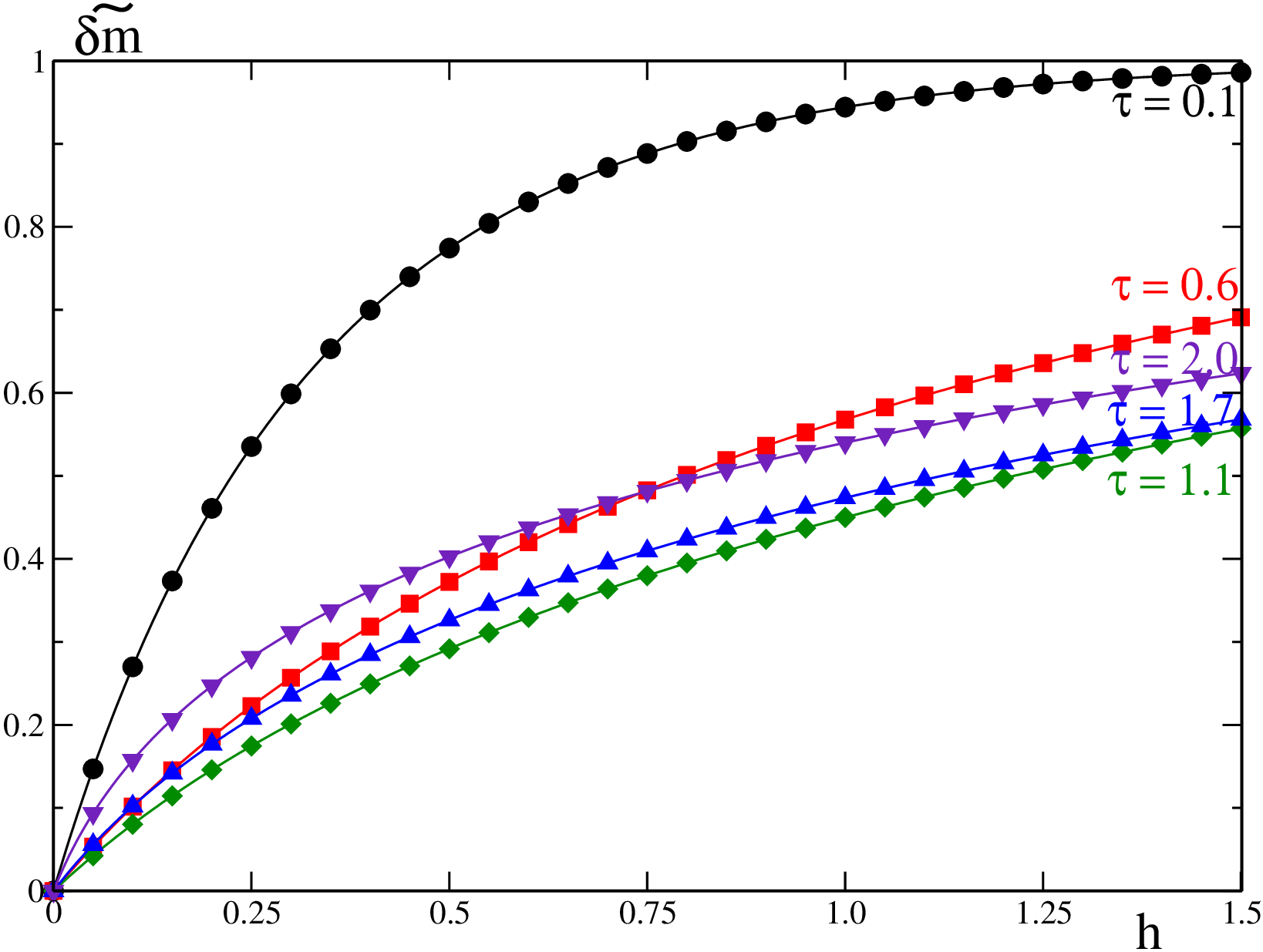} \includegraphics[width=8cm,height=6cm]{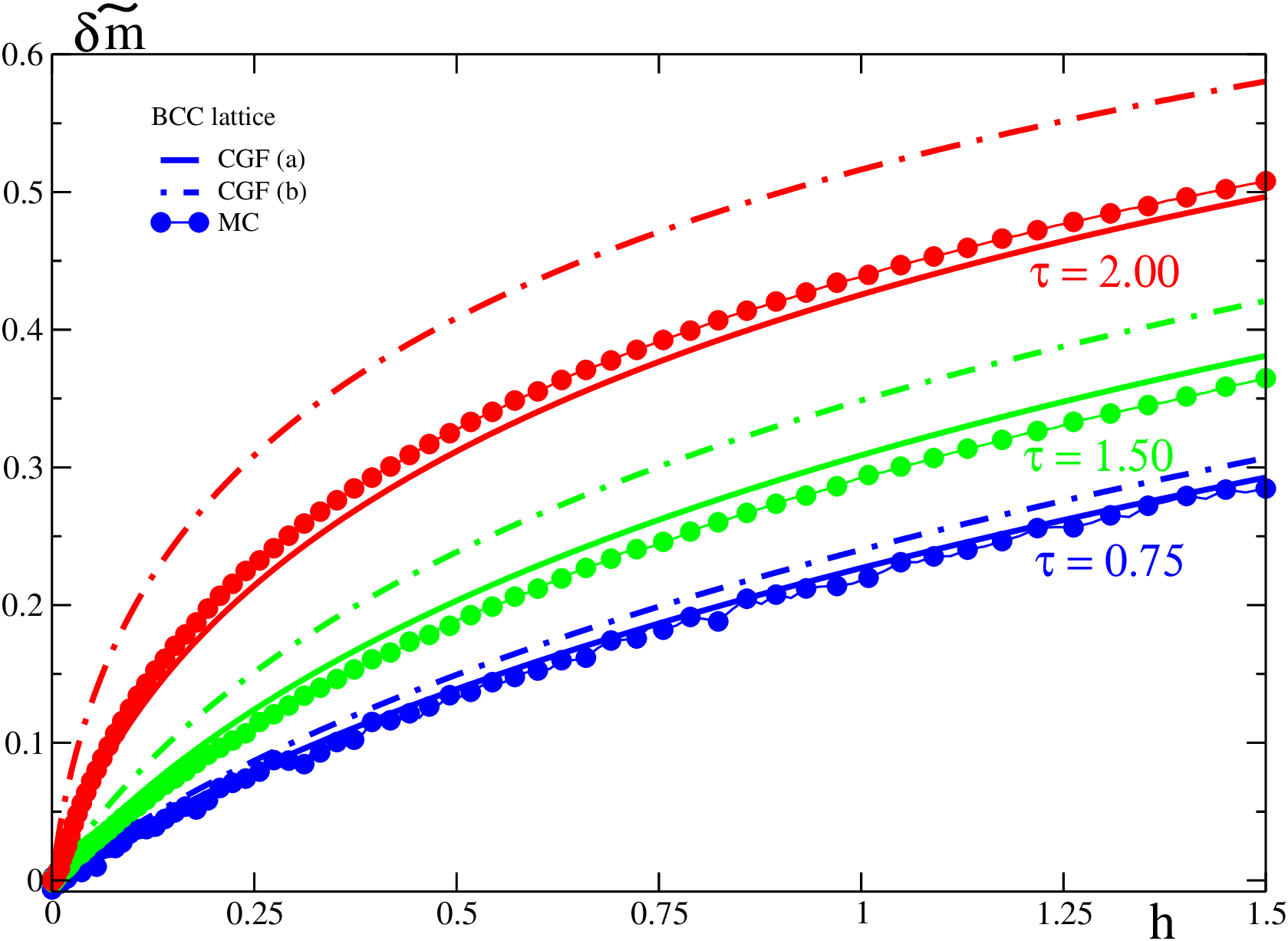} 
\par\end{centering}

\caption{\label{fig:DeltaM(H)Quantum}Field dependence of the relative magnetization
variation $\widetilde{\delta m}$, (left) from QGF with RPA decoupling
and (right) from CGF (Callen (a) and RPA (b) decoupling) and MC.}
\end{figure}

Next, we study the relative magnetization variation

\[
\widetilde{\delta m}\left(T,H\right)=\frac{m(T,H)-m(T,0)}{1-m(T,0)}
\]
 as a function of the applied field for different values of temperature,
without anisotropy. The results are shown in Fig. \ref{fig:DeltaM(H)Quantum}.

In the quantum-mechanical case, we may use the low-temperature asymptote
(\ref{eq:LowTMag}) to get (in the absence of anisotropy)

\begin{align*}
\widetilde{\delta m}\left(T,H\right) & \simeq1-\frac{1}{\zeta\left(3/2\right)}Z{}_{3/2}\left(\frac{g\mu_{B}H}{k_{\mathrm{B}T}}\right).
\end{align*}

This can also be seen within the quantum linear SW theory which renders
exactly the same expression. It is clear from the behavior of $Z_{p}\left(x\right)$
that in the low-temperature regime $\widetilde{\delta m}$ decreases
when the temperature increases.

In the classical case and at low temperature, we use the asymptote
(\ref{eq:CGFLowTAsympt}) and obtain 
\begin{align*}
\widetilde{\delta m}{}^{\mathrm{CGF}} & \simeq1-\left(1+\kappa\right)\psi(1)+\left(1+\kappa\right)\psi(1)\left(\lambda^{\prime}-\psi(1)\right)P_{N}(\psi\left(1\right))\times\tau^{\prime}.
\end{align*}

Here we see that this expression increases when the temperature increases
since $\lambda^{\prime}=1/\left(1+\kappa\right)>\psi\left(1\right)=1/\left(h^{\prime}+\left(1+\kappa\right)\right)$
for any $h^{\prime}>0$.

However, it remains unclear why in the quantum-mechanical case there
is a change of behavior at a particular temperature because it is
difficult to derive an (approximate) analytical expression for the
latter. Indeed, this would at least require to derive the magnetization
asymptote in the critical region in finite magnetic field which, unfortunately,
leads to a rather cumbersome expression. Nevertheless, quantum spin
effects are attenuated at high temperatures and as such the quantum
approach renders the same behavior for the magnetization as the classical
one.

\medskip{}

\subsection{SW spectrum and exchange stiffness at finite temperature}

Now we discuss the exchange stiffness as a function of the magnetization
taking account of nonlinear SW effects. As we have seen, taking account
of these effects (or magnon-magnon interactions), through the various
decoupling schemes, leads to a temperature-dependent dispersion. This
dependence on temperature comes about through the magnetization $m$.
Let us consider, for simplicity, the case with the sole contribution
from exchange coupling. Hence, we may define the SW stiffness $D$
as follows 
\begin{equation}
\hbar\omega_{\mathbf{k}}^{\prime}=D\left(m\right)\left(1-\gamma_{\mathbf{k}}\right),\label{eq:ESDef}
\end{equation}
 assuming that all SW nonlinear effects {[}see the last term in \emph{e.g.}
Eqs. (\ref{eq:NLMagnonDispGF}, \ref{eq:DispCGF}){]} are booked into
the function $D\left(m\right)$. On the other hand, in the absence
of applied field and anisotropy, and for a given decoupling scheme
with the parameter $\alpha$ introduced in Eq. (\ref{eq:SzCallenExpansion})
we deduce from Eq. (\ref{eq:QExpression}) that 
\[
D\left(m\right)=J_{0}^{\prime}mQ^{\prime}\left(\alpha,m\right)
\]
 with 
\[
Q^{\prime}(\alpha,\beta)=1+\frac{\alpha}{\mathcal{N}}\sum_{\mathbf{p}}\frac{\gamma_{\mathbf{p}}}{\beta\hbar\omega_{\mathbf{p}}^{\prime}}
\]
 in the classical limit.

In the general case, as discussed in the previous sections, the dispersion
$\omega_{\mathbf{k}}$ and the magnetization $m$ are solved for by
using the system of coupled equations and then $D\left(m\right)$
is obtained by fitting the curves of $\omega$ as a function of the
wave vector $\mathbf{k}$ in a given direction in Fourier space. Next,
we substitute $\hbar\omega_{\mathbf{k}}^{\prime}=J_{0}^{\prime}mQ^{\prime}\left(1-\gamma_{\mathbf{k}}\right)$
in $Q^{\prime}$ to obtain 
\begin{equation}
Q^{\prime}\left(\alpha,\beta\right)=1+\frac{1}{Q^{\prime}\left(\alpha,\beta\right)}\frac{\alpha/m}{\mathcal{N}}\sum_{\mathbf{p}}\frac{\gamma_{\mathbf{p}}}{1-\gamma_{\mathbf{k^{\prime}}}}\times\tau^{\prime}=1+\frac{\alpha\left(W-1\right)}{mQ^{\prime}\left(\alpha,\beta\right)}\times\tau^{\prime}\label{eq:Qm}
\end{equation}
 where $W$ is the Watson integral for the given lattice.

At low temperature, the magnetization is given by (CGF or CSD)

\[
m\simeq1-\frac{W}{Q^{\prime}\left(\alpha,\beta\right)}\tau^{\prime}
\]
 leading to 
\[
\tau^{\prime}=\left(1-m\right)\frac{Q^{\prime}}{W}.
\]

Then, when this is substituted in Eq. (\ref{eq:Qm}) yields 
\[
Q^{\prime}\left(\alpha,\beta\right)=1+\frac{W-1}{W}\times\frac{\left(1-m\right)\alpha}{m}
\]
 and thereby the spin stiffness $D\left(m\right)$ reads 
\begin{equation}
D\left(m\right)=J_{0}^{\prime}m\left[1+\frac{W-1}{W}\times\frac{\left(1-m\right)\alpha}{m}\right].\label{eq:ExchangeStiff}
\end{equation}

Now, defining $\phi=\alpha\left(m\right)/m$ and $\varsigma\equiv\left(W-1\right)/W$,
we write 
\[
D\left(m\right)=J_{0}^{\prime}m\left[1+\varsigma\phi\left(m\right)\times\left(1-m\right)\right].
\]

Finally, considering the fact that at low temperature $1-m$ is small
so that we may write 
\[
D\left(m\right)\simeq J_{0}^{\prime}m\left[1-\left(1-m\right)\right]^{-\varsigma\phi\left(m\right)}=m^{1-\varsigma\phi\left(m\right)}.
\]

$\alpha\left(m\right)$ and thereby $\phi\left(m\right)$ is given
according to the RPA, Callen's, Copeland and Gersch or Swendsen decoupling
scheme, see Eq. (\ref{eq:SwendsenDecoup}) \emph{et seq}. For the
RPA decoupling, for instance, $\alpha=0$ and thus $D\left(m\right)\sim m$,
as it should. For Callen's decoupling, $\alpha\left(m\right)=m$ leading
to $D\left(m\right)\sim m^{1-\varsigma}$. For a decoupling scheme
with $\alpha\left(m\right)=am+bm^{3}$ we make an expansion around
$m\simeq1$ and obtain $D\left(m\right)\simeq m\left[1-\zeta\left(a+b\right)\left(m-1\right)\right]$.

\begin{figure}[H]

\begin{centering}
\includegraphics[scale=0.4]{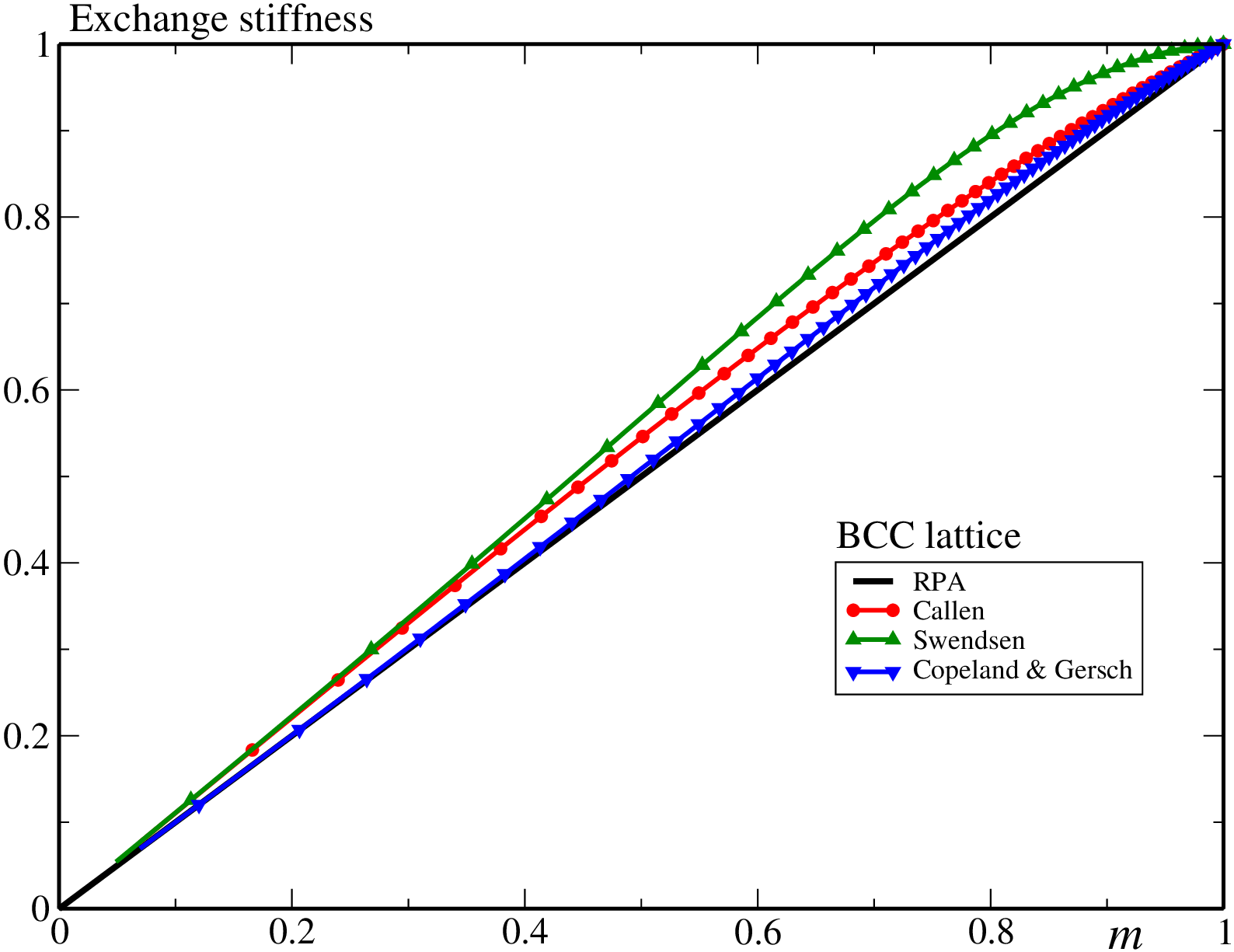} 
\par\end{centering}

\caption{\label{fig:ExchangeStiffness}Exchange stiffness against the magnetization
for different decoupling schemes obtained by CGF. }
\end{figure}

In Fig. \ref{fig:ExchangeStiffness} we plot the exchange stiffness
as obtained numerically from Eq. (\ref{eq:ESDef}) and Eqs. (\ref{eq:NLMagnonDispGF},
\ref{eq:DispCGF}), for the decoupling schemes discussed in Fig. \ref{fig:CGF-decoupling}.

As is seen in Eq. (\ref{eq:ExchangeStiff}) and confirmed by the numerical
results in Fig. \ref{fig:ExchangeStiffness}, the exchange stiffness
depends on the decoupling scheme or the way the spin-spin correlations
are tackled, especially at moderate temperatures. Obviously, the curves
corresponding to the four decoupling schemes merge for $m\sim1$ (very
low temperature) and $m\sim0$\emph{, i.e.} at high temperature where
they exhibit a linear behavior.

\section{\label{sec:Conclusion}Conclusion}

We have established clear connections between the quantum/classical
Green function technique and the classical spectral density method,
and have compared them with the numerical methods of Monte Carlo and
Landau-Lifshitz-Langevin dynamics. We have proposed a unified decoupling
scheme for both anisotropy and exchange contributions, for classical
as well as for quantum spins which allows us to establish a clear
connection between the various methods and to obtain reasonable results
for the magnetization and critical temperature. We have computed the
spin-wave spectrum at finite temperature and inferred the magnetization
as a function of temperature and field and have obtained the exchange
stiffness, for various decoupling schemes. Asymptotic expressions
for the magnetization have been given at low temperature and in the
critical region, both for classical and quantum spins, and the crossover
between them has been established. As far as the (semi-)analytical
methods are concerned, it turns out that the classical Green's function
technique is more straightforward and does not require any \emph{a
priori }assumptions about the the system's spectral density. In particular,
Callen's famous formula for the magnetization is recast in a compact
form using the Brillouin function. This makes it straightforward to
obtain the classical limit leading to the familiar Langevin function
for the magnetization. However, the outcome is still a transcendental
equation involving the spin-wave density, unlike the Langevin form
one obtains from mean-field theory.

In future work, we would like to extend the present calculations and
the Green's function technique to finite-size systems by taking account
of boundary and surface effects, similarly to what has been done in
Refs. \citep{kacgar01physa300, kacgar01epjb}. This should be useful
for studying the dynamics of multi-layered magnetic systems and magnetic
nanostructures.

\section{Acknowledgement}

RB and HK acknowldge financial support from the French National Research
Agency under the program ANR Jeunes-Chercheurs MARVEL. UA and OCF
acknowledge funding by the Spanish Ministry of Science and Innovation
under the grant FIS2010-20979-C02-02.

\appendix

\section{\label{app:QuantumGreenFunctionMethod} Quantum Green Function Method}

In this section we briefly describe the quantum Green function method
(QGM) in the case of an oblique magnetic field.

In order to compute the spin-wave (SW) spectrum and the magnetization,
one deals with the spin fluctuations with respect to the equilibrium
configuration, which has to be determined beforehand. In practice,
one assumes that there exists a net direction of the system's magnetization
denoted by $\mathbf{e}_{3}$

\[
\mathbf{m}=\frac{1}{\mathcal{N}}\sum_{i}\mathbf{S}_{i}\equiv m\,\mathbf{e}_{3}.
\]

We start by passing to the new coordinate system in which the (usually
adopted) $z$ reference direction is now the direction $\mathbf{e}_{3}$.
This amounts to performing a rotation of the original variables $\mathbf{S}_{i}$
to the new ones $\bm{\sigma}_{i}$ around a given axis and at a given
angle depending on $\mathbf{e}_{3}$. Following the standard approach
\citep{pinietal05prb,frokun06pr,Schwiegeretal05pr}, we use the Holstein-Primakov
representation for the new variables $\bm{\sigma}_{i}$ . To rewrite
the Green's functions in the local reference frame, we use a rotation
matrix $\mathcal{R}(\mathbf{e}_{y},\vartheta)$ for the rotation of
an angle $\vartheta$ around the axis $\mathbf{e}_{y}$. So in the
Hamiltonian (\ref{eq:Hamiltonian}) we replace the spin variable $\mathbf{S}_{i}$
by the new one $\bm{\sigma}_{i}$ (with $\left\Vert \bm{\sigma}_{i}\right\Vert =\left\Vert \mathbf{S}_{i}\right\Vert =S$)
using

\begin{align}
\mathbf{S}_{i} & =\mathcal{R}(\mathbf{e}_{y},-\vartheta)\bm{\sigma}_{i}=\cos\vartheta\,\bm{\sigma}_{i}-\sin\vartheta\,(\bm{\sigma}_{i}\times\mathbf{e}_{y})+\left(1-\cos\vartheta\right)(\bm{\sigma}_{i}\cdot\mathbf{e}_{y})\mathbf{e}_{y}.\label{eq:Rotation}
\end{align}

For instance,

\[
S^{z}=\cos\vartheta\sigma_{i}^{z}-\sin\vartheta\sigma_{i}^{x}
\]
 and the Zeeman term ($H_{y}=0$) becomes

\begin{equation}
\mathbf{S}_{i}\cdot\mathbf{H}=\left[H^{x}\cos\vartheta-H^{z}\sin\vartheta\right]\sigma_{i}^{x}+\left[H^{z}\cos\vartheta+H^{x}\sin\vartheta\right]\sigma_{i}^{z}.
\end{equation}
 We also define the rotated field 
\[
\mathbf{H}_{\mathcal{R}}\equiv\mathcal{R}(\mathbf{e}_{y},\vartheta)\mathbf{H}.
\]

The new spin variables satisfy the same algebra as the original spin
variables $\mathbf{S}_{i}$, i.e.,

\begin{eqnarray}
\left[\sigma_{i}^{+},\,\sigma_{j}^{-}\right] & = & 2\delta_{ij}\,\sigma_{i}^{3}\nonumber \\
\left[\sigma_{i}^{3},\,\sigma_{j}^{\mu}\right] & = & \mu\sigma_{i}^{\mu}\delta_{ij},\qquad\mu=\pm.
\end{eqnarray}

Then, rewriting the Hamiltonian (\ref{eq:Hamiltonian}) in the new
variables, we obtain the quadratic form

\begin{eqnarray}
\mathcal{H} & = & -\frac{1}{2}\sum_{i,j=1}^{\mathcal{N}}\sum_{\mu,\nu=+,-,(3)}\sigma_{i}^{\mu}Q_{ij}^{\mu\nu}\sigma_{j}^{\nu}-{\displaystyle {\displaystyle \sum_{i=1}^{\mathcal{N}}}}{\displaystyle {\displaystyle \sum_{\mu=+,-,(3)}}}L^{\mu}\sigma_{i}^{\mu}\label{eq:HamSpinMatrixForm}
\end{eqnarray}
 with the linear coefficients

\begin{equation}
L^{+}=\left(g\mu_{B}\right)\frac{H_{\mathcal{R}}^{-}}{2},\qquad L^{-}=\left(g\mu_{B}\right)\frac{H_{\mathcal{R}}^{+}}{2},\qquad L^{3}=\left(g\mu_{B}\right)H_{\mathcal{R}}^{3}\label{eq:LinearCoefficients}
\end{equation}
 and the quadratic ones

\begin{widetext}

\begin{equation}
\begin{array}{ll}
Q_{ij}^{++}=\frac{K}{2}\sin^{2}\vartheta\,\delta_{ij}=Q_{ij}^{--}, & Q_{ij}^{+-}=\frac{1}{2}\left[J_{ij}+K\sin^{2}\vartheta\delta_{ij}\right]=Q_{ij}^{-+},\\
\\
Q_{ij}^{33}=J_{ij}+2K\cos^{2}\vartheta\delta_{ij} & Q_{ij}^{3+}=Q_{ij}^{+3}=-K\sin\vartheta\cos\vartheta\delta_{ij}=Q_{ij}^{3-}=Q_{ij}^{-3}.
\end{array}\label{eq:QuadraticCoefficients}
\end{equation}
 \end{widetext} These satisfy the symmetry relation $Q_{ij}^{\mu\nu}=Q_{ji}^{\mu\nu}$.

Applying the RPA decoupling to a homogeneous ferromagnet, \emph{i.e.}
with $\left\langle \sigma_{i}^{3}\right\rangle =\left\langle \sigma^{3}\right\rangle $,
we obtain the following (coupled) equations for the relevant GFs after
Fourier transformations with respect to time and space

\begin{widetext} 
\begin{equation}
\left(\begin{array}{lll}
\omega-\mathcal{A}_{\mathbf{k}} & \mathcal{B}_{\mathbf{k}} & 2\mathcal{A}_{\mathbf{k}}^{-}\left(\mathcal{K}_{\sigma}\right)\\
-\mathcal{B}_{\mathbf{k}} & \omega+\mathcal{A}_{\mathbf{k}} & -2\mathcal{A}_{\mathbf{k}}^{+}\left(\mathcal{K}_{\sigma}\right)\\
\mathcal{A}_{\mathbf{k}}^{+}\left(\mathcal{K}_{\sigma}/2\right) & -\mathcal{A}_{\mathbf{k}}^{-}\left(\mathcal{K}_{\sigma}/2\right) & \omega
\end{array}\right)\left(\begin{array}{l}
\mathcal{G}_{\mathbf{k}}^{+-}\\
\mathcal{G}_{\mathbf{k}}^{--}\\
\mathcal{G}_{\mathbf{k}}^{3-}
\end{array}\right)=\left(\begin{array}{l}
2\left\langle \sigma^{3}\right\rangle \\
0\\
0
\end{array}\right)\label{eq:EMSystem}
\end{equation}
 \end{widetext} where 
\begin{equation}
\begin{array}{lll}
\mathcal{A}_{\mathbf{k}} & \equiv & L^{3}+\mathcal{K}_{\sigma}\left\langle \sigma^{3}\right\rangle \left(2\cos^{2}\vartheta-\sin^{2}\vartheta\right)+J_{0}\left\langle \sigma^{3}\right\rangle \left(1-\gamma_{\mathbf{k}}\right),\\
\mathcal{A}_{\mathbf{k}}^{\pm} & \equiv & L^{\pm}-\mathcal{K}_{\sigma}\left\langle \sigma^{3}\right\rangle \sin2\vartheta,\\
\mathcal{B}_{\mathbf{k}} & \equiv & \mathcal{K}_{\sigma}\left\langle \sigma^{3}\right\rangle \sin^{2}\vartheta.
\end{array}\label{eq:EMSystemCoeffs2}
\end{equation}

$J_{0}$ is the $\mathbf{k}=0$ component of the exchange coupling
given by

\begin{equation}
J_{0}\equiv J\left(\text{0}\right)=\sum_{j}J_{ij}=zJ\label{eq:zeroCompJ}
\end{equation}
 with $z$ being the coordination number. If the exchange is isotropic,
we may write 
\begin{align}
J\left(\mathbf{k}\right) & =J\left(-\mathbf{k}\right)=\sum_{j}e^{-i\mathbf{k}\cdot\mathbf{r}_{ij}}J_{ij}=J_{0}\times\frac{1}{z}\sum_{j}e^{-i\mathbf{k}\cdot\mathbf{r}_{ij}}\equiv J_{0}\gamma_{\mathbf{k}}.\label{eq:FTExchCoupling}
\end{align}

For a bcc lattice we have ($z=8$) the unit cell unit vectors

\[
\mathbf{\delta}_{ij}\equiv\frac{a}{2}\left(\pm\mathbf{e}_{_{x}},\pm\mathbf{e}_{y},\pm\mathbf{e}_{z}\right)
\]
 and thereby 
\begin{eqnarray}
\gamma_{\mathbf{k}} & = & \cos\frac{ak_{x}}{2}\cos\frac{ak_{y}}{2}\cos\frac{ak_{z}}{2},\label{eq:BCCDispFct}
\end{eqnarray}
 $a$ being the lattice parameter. For long wavelength excitations
we use $\cos k_{\alpha}\simeq1-\frac{1}{2}k_{\alpha}^{2}$, which
yields 
\begin{eqnarray*}
1-\gamma_{\mathbf{k}} & \simeq & \left(ak\right)^{2}.
\end{eqnarray*}

Note that the EM for $\mathcal{G}_{ij}^{z-}$, that is the last equation
in the system (\ref{eq:EMSystem}), provides the equilibrium configuration.
Near equilibrium, the net magnetic moment $\mathbf{m}=\frac{1}{\mathcal{N}}\sum_{i}\mathbf{\sigma}_{i}$
does not change much, which means that $d\mathbf{m}/dt\simeq0$. In
quantum mechanics, this implies that the total magnetic moment along
the equilibrium direction commutes with the Hamiltonian, or in other
words, the projection of the total magnetic moment along the equilibrium
direction is conserved, that is

\begin{equation}
i\frac{d}{dt}\left(\frac{1}{\mathcal{N}}\sum_{i}\sigma_{i}^{3}\right)=\left[\frac{1}{\mathcal{N}}\sum_{i}\sigma_{i}^{3},\mathcal{H}\right]=0.\label{eq:EquilibriumCond}
\end{equation}

On the other hand, on the same level of approximation as that used
to obtain the system of EM (\ref{eq:EMSystem}), the commutator above
reads

\begin{align}
\left[\frac{1}{\mathcal{N}}\sum_{i}\sigma_{i}^{3},\mathcal{H}\right] & \simeq\left[\left(g\mu_{B}\right)\left(H^{z}\sin\vartheta-H^{x}\cos\vartheta\right)+\mathcal{K}\left\langle \sigma^{3}\right\rangle \sin2\vartheta\right]\times\frac{1}{\mathcal{N}}\sum_{i}\left(i\sigma_{i}^{y}\right),\label{eq:ApproxCommutator}
\end{align}
 which, if set to zero according to (\ref{eq:EquilibriumCond}), leads
to the equilibrium condition 
\begin{equation}
\left(h^{z}\sin\vartheta-h^{x}\cos\vartheta\right)+\mathcal{K}_{\sigma}\left\langle \sigma^{3}\right\rangle \sin2\vartheta=0.\label{eq:EquilCondition}
\end{equation}
 Hence, the GF $\mathcal{G}_{\mathbf{k}}^{z-}(\omega)$ is eliminated
from the system (\ref{eq:EMSystem}) and thereby the latter simplifies
into the following system of two coupled equations 
\begin{equation}
\left\{ \begin{array}{ccc}
\left(\omega-\mathcal{A}_{\mathbf{k}}\right)\mathcal{G}_{\mathbf{k}}^{+-}+\mathcal{B}_{\mathbf{k}}\mathcal{G}_{\mathbf{k}}^{--} & = & 2\left\langle \sigma^{3}\right\rangle \\
\\
-\mathcal{B}_{\mathbf{k}}\mathcal{G}_{\mathbf{k}}^{+-}+\left(\omega+\mathcal{A}_{\mathbf{k}}\right)\mathcal{G}_{\mathbf{k}}^{--} & = & 0.
\end{array}\right.\label{eq:GFSystem}
\end{equation}

\section{\label{app:CSDMExchAnisDecoup}Decoupling of exchange contributions
within CSD}

Following the procedure described in section \ref{sub:CSD}, before
Eq. (\ref{eq:DecouplingExchangeCSD}), we obtain 
\begin{align*}
\frac{1}{\mathcal{N}}\sum_{\mathbf{q}}J_{\mathbf{\mathbf{q}}}^{\prime}\begin{array}{c}
\left\langle \left\{ s_{\mathbf{k}-\mathbf{q}}^{z}(\tau)s_{\mathbf{q}}^{+}(\tau);s_{-\mathbf{k}}^{-}(0)\right\} \right\rangle _{\omega}\end{array} & \simeq mJ_{\mathbf{k}}^{\prime}\left\langle \left\langle s_{\mathbf{k}}^{+}(\tau);s_{-\mathbf{k}}^{-}\right\rangle \right\rangle _{\omega}\\
 & -\frac{m}{\mathcal{N}^{2}}\sum_{\mathbf{p},\mathbf{q}}J_{\mathbf{\mathbf{q}}}^{\prime}\frac{\left\langle s_{\mathbf{p}}^{+}s_{\mathbf{q}}^{-}\right\rangle }{2}\left\langle \left\langle s_{\mathbf{k}-\mathbf{p}-\mathbf{q}}^{+}(\tau);s_{-\mathbf{k}}^{-}\right\rangle \right\rangle _{\omega}.
\end{align*}

Similarly, for the second contribution we get 
\begin{align*}
\frac{1}{\mathcal{N}}\sum_{\mathbf{q}}J_{\mathbf{\mathbf{q}}}^{\prime}\left\langle \left\{ s_{\mathbf{q}}^{z}(\tau)s_{\mathbf{k}-\mathbf{q}}^{+}(\tau);s_{-\mathbf{k}}^{-}(0)\right\} \right\rangle _{\omega} & \simeq mJ_{\mathbf{\mathbf{0}}}^{\prime}\left\langle \left\langle s_{\mathbf{k}}^{+}(\tau);s_{-\mathbf{k}}^{-}\right\rangle \right\rangle _{\omega}\\
 & -\frac{m}{\mathcal{N}^{2}}\sum_{\mathbf{p},\mathbf{q}}J_{\mathbf{\mathbf{q}}}^{\prime}\frac{\left\langle s_{\mathbf{p}}^{+}s_{\mathbf{k}-\mathbf{q}}^{-}\right\rangle }{2}\left\langle \left\langle s_{\mathbf{q}-\mathbf{p}}^{+}(\tau);s_{-\mathbf{k}}^{-}\right\rangle \right\rangle _{\omega}.
\end{align*}

Now, using the two moment equations 
\begin{equation}
\left\{ \begin{array}{lll}
\intop_{-\infty}^{\infty}\frac{d\omega}{2\pi}\Lambda_{\mathbf{k}}\left(\omega\right)=i\left\langle \left\{ S_{\mathbf{k}}^{+},S_{-\mathbf{k}}^{-}\right\} \right\rangle ,\\
\\
\intop_{-\infty}^{\infty}\frac{d\omega}{2\pi}\omega\Lambda_{\mathbf{k}}\left(\omega\right)=-\left\langle \left\{ \left\{ S_{\mathbf{k}}^{+},\mathcal{H}\right\} ,S_{-\mathbf{k}}^{-}\right\} \right\rangle .
\end{array}\right.\label{eq:TwoME}
\end{equation}
 with the spectral density, see Ref. \cite{cavalloetal05review} 
\begin{equation}
\Lambda_{\mathbf{k}}\left(\omega\right)=\left\langle \left\langle s_{\mathbf{k}}^{+}(\tau);s_{-\mathbf{k}}^{-}\right\rangle \right\rangle _{\omega}=i\left\langle \left\{ s_{\mathbf{k}}^{+}\left(\tau\right),s_{-\mathbf{k}}^{-}\left(0\right)\right\} \right\rangle _{\omega}=2\pi\mathcal{N}m\delta\left(\omega-\omega_{\mathbf{k}}\right)\label{eq:1deltaSD}
\end{equation}
 we integrate over $\omega$ and obtain for the first contribution
\begin{align*}
\intop_{-\infty}^{\infty}\frac{d\omega}{2\pi}\frac{1}{\mathcal{N}}\sum_{\mathbf{q}}J_{\mathbf{\mathbf{q}}}^{\prime}\begin{array}{c}
\left\langle \left\{ s_{\mathbf{k}-\mathbf{q}}^{z}(\tau)s_{\mathbf{q}}^{+}(\tau);s_{-\mathbf{k}}^{-}(0)\right\} \right\rangle _{\omega}\end{array} & \simeq2\mathcal{N}m^{2}J_{\mathbf{k}}^{\prime}-\frac{2m}{\mathcal{N}}\sum_{\mathbf{q}}J_{\mathbf{\mathbf{q}}}^{\prime}\left\langle s_{-\mathbf{q}}^{+}s_{\mathbf{q}}^{-}\right\rangle 
\end{align*}
 and for the second 
\begin{align*}
\intop_{-\infty}^{\infty}\frac{d\omega}{2\pi}\frac{1}{\mathcal{N}}\sum_{\mathbf{q}}J_{\mathbf{\mathbf{q}}}^{\prime}\left\langle \left\{ s_{\mathbf{q}}^{z}(\tau)s_{\mathbf{k}-\mathbf{q}}^{+}(\tau);s_{-\mathbf{k}}^{-}(0)\right\} \right\rangle _{\omega} & \simeq2\mathcal{N}m^{2}J_{\mathbf{\mathbf{0}}}^{\prime}-\frac{m^{2}}{\mathcal{N}}\sum_{\mathbf{q}}J_{\mathbf{\mathbf{q}}}^{\prime}\left\langle s_{\mathbf{q}-\mathbf{k}}^{+}s_{\mathbf{k}-\mathbf{q}}^{-}\right\rangle .
\end{align*}

Then, subtracting the second contribution from the first yields 
\begin{align*}
\intop_{-\infty}^{\infty}\frac{d\omega}{2\pi}\frac{1}{\mathcal{N}}\sum_{\mathbf{q}}J_{\mathbf{\mathbf{q}}}^{\prime}\left[\begin{array}{c}
\left\langle \left\{ s_{\mathbf{k}-\mathbf{q}}^{z}(\tau)s_{\mathbf{q}}^{+}(\tau);s_{-\mathbf{k}}^{-}(0)\right\} \right\rangle _{\omega}\\
-\left\langle \left\{ s_{\mathbf{q}}^{z}(\tau)s_{\mathbf{k}-\mathbf{q}}^{+}(\tau);s_{-\mathbf{k}}^{-}(0)\right\} \right\rangle _{\omega}
\end{array}\right] & \simeq2\mathcal{N}m^{2}\left(J_{\mathbf{k}}^{\prime}-J_{\mathbf{\mathbf{0}}}^{\prime}\right)-\frac{m^{2}}{\mathcal{N}}\sum_{\mathbf{q}}\left(J_{\mathbf{\mathbf{q}}}^{\prime}-J_{\mathbf{\mathbf{k}}-\mathbf{\mathbf{q}}}^{\prime}\right)\left\langle s_{-\mathbf{q}}^{+}s_{\mathbf{q}}^{-}\right\rangle .
\end{align*}

On the other hand, using the zero-moment equation, we get 
\begin{align*}
\intop_{-\infty}^{\infty}\frac{d\omega}{2\pi}\frac{1}{\mathcal{N}}\sum_{\mathbf{q}}J_{\mathbf{\mathbf{q}}}^{\prime}\left[\begin{array}{c}
\left\langle \left\{ s_{\mathbf{k}-\mathbf{q}}^{z}(\tau)s_{\mathbf{q}}^{+}(\tau);s_{-\mathbf{k}}^{-}(0)\right\} \right\rangle _{\omega}\\
-\left\langle \left\{ s_{\mathbf{q}}^{z}(\tau)s_{\mathbf{k}-\mathbf{q}}^{+}(\tau);s_{-\mathbf{k}}^{-}(0)\right\} \right\rangle _{\omega}
\end{array}\right] & =-\frac{1}{\mathcal{N}}\sum_{\mathbf{q}}\left(J_{\mathbf{\mathbf{q}}}^{\prime}-J_{\mathbf{\mathbf{k}}-\mathbf{\mathbf{q}}}^{\prime}\right)\left[2\left\langle s_{\mathbf{q}}^{z}s_{-\mathbf{q}}^{z}\right\rangle +\left\langle s_{-\mathbf{q}}^{-}s_{\mathbf{q}}^{+}\right\rangle \right].
\end{align*}

Consequently, we have the equation 
\[
2\mathcal{N}m^{2}\left(J_{\mathbf{\mathbf{0}}}^{\prime}-J_{\mathbf{k}}^{\prime}\right)+\frac{m^{2}}{\mathcal{N}}\sum_{\mathbf{q}}\left(J_{\mathbf{\mathbf{q}}}^{\prime}-J_{\mathbf{\mathbf{k}}-\mathbf{\mathbf{q}}}^{\prime}\right)\left\langle s_{-\mathbf{q}}^{+}s_{\mathbf{q}}^{-}\right\rangle \simeq\frac{1}{\mathcal{N}}\sum_{\mathbf{q}}\left(J_{\mathbf{\mathbf{q}}}^{\prime}-J_{\mathbf{\mathbf{k}}-\mathbf{\mathbf{q}}}^{\prime}\right)\left[2\left\langle s_{\mathbf{q}}^{z}s_{-\mathbf{q}}^{z}\right\rangle +\left\langle s_{-\mathbf{q}}^{-}s_{\mathbf{q}}^{+}\right\rangle \right]
\]
 which leads to 
\[
\sum_{\mathbf{q}}\left(J_{\mathbf{\mathbf{q}}}^{\prime}-J_{\mathbf{\mathbf{k}}-\mathbf{\mathbf{q}}}^{\prime}\right)\left\langle s_{\mathbf{q}}^{z}s_{-\mathbf{q}}^{z}\right\rangle \simeq\mathcal{N}^{2}m^{2}\left(J_{\mathbf{\mathbf{0}}}^{\prime}-J_{\mathbf{k}}^{\prime}\right)+\frac{1}{2}m^{2}\sum_{\mathbf{q}}\left(J_{\mathbf{\mathbf{q}}}^{\prime}-J_{\mathbf{\mathbf{k}}-\mathbf{\mathbf{q}}}^{\prime}\right)\left\langle s_{-\mathbf{q}}^{+}s_{\mathbf{q}}^{-}\right\rangle -\frac{1}{2}\sum_{\mathbf{q}}\left(J_{\mathbf{\mathbf{q}}}^{\prime}-J_{\mathbf{\mathbf{k}}-\mathbf{\mathbf{q}}}^{\prime}\right)\left\langle s_{-\mathbf{q}}^{-}s_{\mathbf{q}}^{+}\right\rangle .
\]

One can can easily check that $\left\langle s_{-\mathbf{q}}^{+}s_{\mathbf{q}}^{-}\right\rangle =\left\langle s_{-\mathbf{q}}^{-}s_{\mathbf{q}}^{+}\right\rangle $
and thereby one obtains

\[
\sum_{\mathbf{q}}\left(J_{\mathbf{\mathbf{q}}}^{\prime}-J_{\mathbf{\mathbf{k}}-\mathbf{\mathbf{q}}}^{\prime}\right)\left\langle s_{\mathbf{q}}^{z}s_{-\mathbf{q}}^{z}\right\rangle \simeq\sum_{\mathbf{q}}\left(J_{\mathbf{\mathbf{q}}}^{\prime}-J_{\mathbf{\mathbf{k}}-\mathbf{\mathbf{q}}}^{\prime}\right)\left[\mathcal{N}^{2}m^{2}\Delta\left(\mathbf{q}\right)-\frac{1}{2}\left(1-m^{2}\right)\left\langle s_{\mathbf{q}}^{+}s_{-\mathbf{q}}^{-}\right\rangle \right].
\]
 This may also be recast into the form (\ref{eq:DecouplingExchangeCSD})
which can be more easily compared to RPA.

\section{\label{app:TC}The critical temperature via different approaches.}

Within the QGF approach and using parameter $\phi$ for exchange decoupling,
the Curie temperature can be calculated from Eq. (\ref{eq:MagCGFNearTCPhi})
by setting $\left\langle \sigma^{3}\right\rangle \simeq0$ at $\tau=\tau_{c}$.
This leads to {[}see Eq. (\ref{eq:NearTcDispParams}) for notation{]}

\[
\tau_{c}=\frac{S(S+1)}{3}\frac{\kappa+Q(\phi,\tau_{c})}{P(\Lambda)}.
\]

In the absence of anisotropy, which is negligible near $T_{c}$, we
obtain

\begin{equation}
\tau_{c}^{\mathrm{QGF}}=\frac{S(S+1)}{3}\frac{Q_{\mathrm{exch}}(\phi,\tau_{c}^{\mathrm{QGF}})}{W}\rightarrow\tau_{c}^{\mathrm{QGF}}=\frac{S\left(S+1\right)}{3W}\left[1+\frac{\phi}{3}\left(1+\frac{1}{S}\right)\left(1-\frac{1}{W}\right)\right].\label{eq:QGFTc}
\end{equation}

In CGF (or in CSD where we obtain the same result), we similarly use
the high-temperature asymptote (\ref{eq:CGFHightTAsymptQ}) and obtain

\begin{equation}
\tau_{c}^{\mathrm{CGF}}=S^{2}\frac{Q_{\mathrm{exch}}^{\prime}(\phi,\tau_{c}^{\mathrm{CGF}})}{3W}\rightarrow\tau_{c}^{\mathrm{CGF}}=\frac{S^{2}}{3W}\left[1+\frac{\phi}{3}\left(1-\frac{1}{W}\right)\right].\label{eq:CGFTc}
\end{equation}

We remark in passing that this is also the result that one obtains
within the spherical model, in the isotropic case \citep{kacgar01physa300},
\emph{i.e.} $\kappa=0$, and for a RPA decoupling $\phi=0$, which
yields

\begin{equation}
\tau_{c}=\frac{1}{W}\frac{S^{2}}{3}.\label{eq:SMTc-1}
\end{equation}

On the other hand, from the MFT magnetization (\ref{eq:LangevinMag})
one obtains the Curie temperature (for $H=0$ and $\kappa=0$)

\begin{equation}
\tau_{C}^{\mathrm{MFT}}=\frac{J_{0}S^{2}}{3}.\label{eq:MFTTc-1}
\end{equation}

Note that contrary to the MFT result (\ref{eq:MFTTc-1}), the expression
(\ref{eq:CGFTc}) for $\tau_{C}$, as obtained from the GF in the
classical limit, or Eq. (\ref{eq:SMTc-1}) from the isotropic spherical
model, depends on the lattice and on the SW dispersion via the Watson
integral $W$. Moreover, as mentioned earlier, we can relate MFT to
SWT by assuming that all excitations are degenerate and by ignoring
spin fluctuations. More precisely, this amounts to dropping the terms
that are responsible for the propagation of the SWs (or magnons) through
the lattice. This can be done by dropping the propagation function
$\gamma_{\mathbf{k}}$ from all SWT expressions. Hence, the MFT result
(\ref{eq:MFTTc-1}) can be obtained from the classical limit of the
GF result (\ref{eq:CGFTc}) by formally setting $\gamma_{\mathbf{k}}=0$
in the lattice integral $W$ (leading to $W=1$) and taking $\phi=0$.

In Table \ref{tab:TC} we collect the values of $\tau_{C}$ estimated
by the different approaches in the isotropic case. First, we remark
that the values obtained within quantum-mechanical approaches are
higher than the classical ones. Indeed, comparing for instance Eqs.
(\ref{eq:QGFTc}) and (\ref{eq:CGFTc}) we see that for small spin
values the difference in $\tau_{c}$, due to the contribution $S$
in $S\left(S+1\right)=S^{2}+S$, is non negligible. This is no surprise
because this decoupling scheme, unlike RPA, accounts for magnon-magnon
interactions whose role becomes crucial in the vicinity of the critical
temperature. Second, there is a perfect agreement between the two
classical methods CGF and CSD. As discussed earlier, this shows that
given that i) CGF renders the same results as CSD and ii) that CGF
does not require any assumptions about the spectral function, it might
be preferable to use the CGF method.

\begin{table}[H]
\begin{centering}
\begin{tabular}{|c|c|c|c|c|c|}
\hline 
Method  & $\begin{array}{ccc}
 & \mbox{QGF} & \mathrm{}\\
\mathrm{(a)} & \mathrm{(b)} & \mathrm{(c)}
\end{array}$  & $\begin{array}{ccc}
 & \mbox{CGF} & \mathrm{}\\
\mathrm{(a)} & \mathrm{(b)} & \mathrm{(c)}
\end{array}$  & $\begin{array}{ccc}
 & \mbox{CSD} & \mathrm{}\\
\mathrm{(a)} & \mathrm{(b)} & \mathrm{(c)}
\end{array}$  & MFT  & MC\tabularnewline
\hline 
\hline 
$\tau_{\mathrm{C}}/S^{2}$ (K)  & $\begin{array}{cccc}
0.335 & 0.380 &  & 0.354\end{array}$  & $\begin{array}{ccc}
0.240 & 0.262 & 0.262\end{array}$  & $\begin{array}{ccc}
0.240 & 0.262 & 0.262\end{array}$  & $0.333$  & $0.268$\tabularnewline
\hline 
\end{tabular}
\par\end{centering}

\caption{\label{tab:TC}Reduced Curie temperature $\tau_{\mathrm{C}}/S^{2}\equiv k_{\mathrm{B}}T_{\mathrm{C}}/S^{2}J_{0}$
for a bcc lattice with Fe parameters and $S=5/2$. (a) stands for
the RPA or Copeland-Gersch, (b) for Callen, and (c) for Swenden's
decoupling schemes used for the exchange contributions.}
\end{table}


\expandafter\ifx\csname natexlab\endcsname\relax\global\long\def\natexlab#1{#1}
 \fi \expandafter\ifx\csname bibnamefont\endcsname\relax \global\long\def\bibnamefont#1{#1}
 \fi \expandafter\ifx\csname bibfnamefont\endcsname\relax \global\long\def\bibfnamefont#1{#1}
 \fi \expandafter\ifx\csname citenamefont\endcsname\relax \global\long\def\citenamefont#1{#1}
 \fi \expandafter\ifx\csname url\endcsname\relax \global\long\def\url#1{\texttt{#1}}
 \fi \expandafter\ifx\csname urlprefix\endcsname\relax\global\long\def\urlprefix{URL }
 \fi \providecommand{\bibinfo}[2]{#2} \providecommand{\eprint}[2][]{\url{#2}}

\end{document}